\documentclass[11pt]{article}
\usepackage{geometry}
\usepackage{amsmath}
\usepackage{amssymb}
\usepackage{latexsym}
\usepackage{epsfig}

\usepackage[vcentermath]{youngtab}

\usepackage{amsmath,amssymb,amsbsy}
\usepackage{latexsym,layout,amsfonts,amssymb,fontenc,xcolor,amsthm,multirow,amsfonts,bm,textcomp}
\usepackage{hyperref}
\usepackage{empheq}
\usepackage{accents}

\newsavebox{\uuunit}
\sbox{\uuunit}
    {\setlength{\unitlength}{0.825em}
     \begin{picture}(0.6,0.7)
        \thinlines
        \put(0,0){\line(1,0){0.5}}
        \put(0.15,0){\line(0,1){0.7}}
        \put(0.35,0){\line(0,1){0.8}}
       \multiput(0.3,0.8)(-0.04,-0.02){12}{\rule{0.5pt}{0.5pt}}
     \end {picture}}

\def\equationautorefname~#1\null{eq.~(#1)\null
}

\hypersetup{
    bookmarks=false,         
    unicode=false,          
    pdftoolbar=true,        
    pdfmenubar=true,        
    pdffitwindow=false, 
    pdfstartview={FitH},    
    pdftitle={Tadpole conditions for exotic branes},    
    pdfauthor={Davide M. Lombardo, Fabio Riccioni, Stefano Risoli},     
    pdfsubject={},   
    pdfnewwindow=true,      
    colorlinks=false,       
    linkcolor=blue,          
    citecolor=blue,        
    filecolor=blue,      
    urlcolor=blue           
    linkbordercolor={blue},
    citebordercolor={purple},
    urlbordercolor={gray}
}

\begin{document}

\begin{titlepage}
\begin{center}


\vskip 2cm

{\Large \bf Non-geometric fluxes\\
\vskip .5cm
\& \\
\vskip .5cm
tadpole conditions for exotic branes}

\vskip 2.5cm

{\bf  Davide M. Lombardo\,$^1$, Fabio Riccioni\,$^2$ and Stefano Risoli\,$^{1,2}$}

\vskip 35pt

{\em $^1$ \hskip -.1truecm  Dipartimento di Fisica, Universit\`a di Roma ``La Sapienza'',\\ Piazzale Aldo Moro 2, 00185 Roma, Italy
 \vskip 5pt }

\vskip 15pt

{\em $^2$ \hskip -.1truecm
 INFN Sezione di Roma,   Dipartimento di Fisica, Universit\`a di Roma ``La Sapienza'',\\ Piazzale Aldo Moro 2, 00185 Roma, Italy
 \vskip 25pt }

{email addresses: {\tt Lombardo.1651528@studenti.uniroma1.it},  {\tt Fabio.Riccioni@roma1.infn.it}, {\tt Stefano.Risoli@roma1.infn.it}  \vskip 25pt}

\end{center}

\vskip .5cm

\begin{center} {\bf ABSTRACT}\\[3ex]
\end{center}

We extend the $\mathcal{P}$-flux analysis carried out recently on the  $T^6/[\mathbb{Z}_2 \times \mathbb{Z}_2 ]$ type-II orientifold model to include all the possible non-geometric fluxes. By deriving universal T-duality rules for all the fluxes, we are able to write down a complete expression for the superpotential for both the IIB and IIA theories. By exploiting the universal T-duality rules that apply to all the branes in string theory, 
we then identify all the exotic branes that can be consistently included to cancel  the tadpoles induced by the fluxes. Finally, we derive the representations of these branes with respect to the $SL(2,\mathbb{Z})^7$ duality symmetry of the model.

\end{titlepage}

\newpage
\setcounter{page}{1} \tableofcontents

\vskip 2truecm

\setcounter{page}{1} \numberwithin{equation}{section}

\section{Introduction}

The general lesson that one learns from T-duality is that in string theory one has to generalise the concept of geometry. In particular, in the context of flux compactifications, T-duality  implies the presence of non-geometric fluxes, that are quantities that cannot be obtained in terms of the fields of the higher-dimensional supergravity theory. In the case of torus reductions with fluxes turned on, while the  RR fluxes of the IIA and IIB theory are mapped into each other by T-duality in the standard way as 
\begin{equation}
\mathcal{F}_{a b_1 ...b_p} \overset{{\rm T}_a}{\longleftrightarrow} \mathcal{F}_{b_1 ...b_p} \quad , \label{TdualityruleRRfluxes}
\end{equation}
the NS 3-form flux $\mathcal{H}_{abc}$ and the metric flux $f_{ab}^{c}$ transform into the non-geometric $\mathcal{Q}_{a}^{bc}$ and $\mathcal{R}^{abc}$ fluxes according to the chain of rules \cite{Shelton:2005cf} (see also \cite{Wecht:2007wu})
\begin{equation}
\mathcal{H}_{abc} \overset{{\rm T}_c}{\longleftrightarrow} -f_{ab}^c \overset{{\rm T}_b}{\longleftrightarrow} -\mathcal{Q}_a^{bc} \overset{{\rm T}_a}{\longleftrightarrow} \mathcal{R}^{abc}  \quad . \label{TdualityruleNSfluxes}
\end{equation}

In this paper we will focus on  the IIB/O3 and the dual IIA/O6   $T^6/[\mathbb{Z}_2 \times \mathbb{Z}_2 ]$ orientifold models \cite{orientifolds} with generalised fluxes turned on.  In the IIB/O3-orientifold picture, the only geometric fluxes that can be introduced are $\mathcal{F}_3$ and $\mathcal{H}_3$. The dual IIA/O6-orientifold is obtained by performing three specific T-dualities, which corresponds to mirror symmetry for this particular orbifold \cite{Strominger:1996it}.\footnote{A general analysis of the type II flux solutions which can be generated by T-dualizing the factorized $T^2 \times T^2 \times T^2$ torus in the geometric case has been carried out in \cite{Maxfield:2013wka}.}
This maps the $\mathcal{H}_3$ flux of IIB to both geometric and non-geometric fluxes in IIA. If one then includes 
all the allowed fluxes in the IIA picture, this corresponds  in the IIB picture to also including the $\mathcal{Q}$ flux.\footnote{The full list of RR, NS and $\mathcal{P}$ fluxes that can be turned on in the model, and how they are mapped by mirror symmetry, is given in Table \ref{RRNSNSPfluxestable}.} The resulting ${\cal N}=1$ superpotential was originally derived in \cite{Shelton:2005cf}.

In the IIB theory one can also include the flux $\mathcal{P}_a^{bc}$ which is the S-dual of $\mathcal{Q}_a^{bc}$, and derive the way this flux enters the ${\cal N}=1$ superpotential \cite{Aldazabal:2006up}.  
In \cite{Bergshoeff:2015cba} it was shown that the fluxes that are related by T-duality to $\mathcal{P}_a^{bc}$  in any dimension are  $\mathcal{P}_a^{b_1 ..b_p}$ and $\mathcal{P}^{a,b_1 ...b_p}$, where $p$ is even in IIB and odd in IIA and the $b$ indices are completely antisymmetrised.\footnote{The $\mathcal{P}$ fluxes with all upstairs indices belong to mixed-symmetry irreducible representations.}  In particular, T-duality acts according to the rule \cite{Lombardo:2016swq}
\begin{align}
& \mathcal{P}_a^{b_1 ... b_p} \   \overset{{\rm T}_a}{\longleftrightarrow} \ \mathcal{P}^{a, b_1 ... b_p a} \nonumber \\
& \mathcal{P}_a^{b_1 ... b_p} \  \overset{{\rm T}_{b_p}}{\longleftrightarrow} \ \mathcal{P}_{a}^{b_1 ... b_{p-1}} \label{TdualityrulePfluxes} \\
 & \mathcal{P}^{a ,b_1 ... b_p} \   \overset{{\rm T}_{b_p}}{\longleftrightarrow} \ \mathcal{P}^{a , b_1 ... b_{p-1}} \quad ,\nonumber 
\end{align}
and the fluxes are such that if $a$ is upstairs it coincides with one of the $b$ indices, while if it is downstairs it differs from all the $b$ indices.
Using this rule, one can determine for the $T^6/[\mathbb{Z}_2 \times \mathbb{Z}_2 ]$ orientifold which fluxes can be included in both the IIB and IIA theory following precisely the same method  used for the NS fluxes: one first maps the flux $\mathcal{P}_a^{bc}$ of IIB to the dual IIA theory by performing three T-dualities, then extends this to include all the allowed IIA fluxes, and finally one maps this back to IIB. It turns out that in IIB one has to include the additional flux $\mathcal{P}^{a,b_1 ...b_4}$. 
The expression for the resulting superpotential of each theory is mapped to the one of the other theory under mirror symmetry \cite{Lombardo:2016swq}, and  in the IIB case it  coincides with the one derived in \cite{Aldazabal:2010ef} as far as the $\mathcal{P}$ fluxes are concerned. 

In string theory, fluxes cannot be turned on arbitrarily, because the presence of Chern-Simons terms in the ten-dimensional type-II supergravity actions implies that fluxes along internal cycles generate effective charges for specific potentials, that for consistency have to be cancelled. In particular, if the potential is projected out in the orientifold model, this implies that the charge induced by the fluxes has to vanish, resulting in a quadratic constraint which is the Bianchi identity for the magnetic dual of the potential. Otherwise, one has to impose a tadpole condition, which is the condition that the charge induced by the fluxes has  to be cancelled by including a suitable number of branes. Specifically, the geometric fluxes give rise to non-trivial NS Bianchi identities, but also to a D3 tadpole in IIB, originating from  the term $\mathcal{H}_3 \wedge \mathcal{F}_3$, and  a D6 tadpole in IIA from $\mathcal{H}_3 \mathcal{F}_0 + f \cdot \mathcal{F}_2$. When also the non-geometric  $\mathcal{Q}$, $\mathcal{R}$ and $\mathcal{P}_a^{bc}$  fluxes are included, this gives further contributions to the NS Bianchi identities, but also generates additional tadpoles \cite{Shelton:2005cf,Aldazabal:2006up}.
In particular, in IIB the term $\mathcal{Q}_{[a}^{bc} \mathcal{F}_{d]bc}$ generates a tadpole for the D7-brane, and  by S-duality this is mapped to the  term $\mathcal{P}_{[a}^{bc }\mathcal{H}_{d]bc}$ which generates a tadpole for the S-dual of the D7-brane \cite{Aldazabal:2006up}. 

The fact that the flux $\mathcal{P}_a^{bc}$ of the IIB theory transforms under T-duality as in eq. \eqref{TdualityrulePfluxes} implies that one can map to the IIA theory the tadpole condition for the S-dual of the D7-brane  by performing three T-dualities. This results in a tadpole condition  for an exotic brane \cite{Lombardo:2016swq}, which is an object that in  the higher-dimensional theory  is  a generalised KK-monopole, {\it i.e.} an object well-defined only in the presence of isometries \cite{Elitzur:1997zn,Obers:1998fb,LozanoTellechea:2000mc,deBoer:2010ud,deBoer:2012ma,Chatzistavrakidis:2013jqa}. In \cite{Bergshoeff:2011zk,Bergshoeff:2011ee,Bergshoeff:2011se,Bergshoeff:2012ex} it was shown that exotic branes are associated to specific components of ten-dimensional mixed-symmetry potentials. In particular, given a  ten-dimensional mixed-symmetry potential $A_{p,q,r,..}$  in a representation such that $p,q,r, ...$ (with $p\ \geq q \geq r ...$) denote the length of each column of its Young Tableau, this corresponds to a brane if some of the  indices $p$  are isometries and contain all the indices $q$, which themselves contain all the indices $r$ and so on. One can then classify all the mixed-symmetry potentials that give rise to branes in lower dimensions in terms of the non-positive integer number $\alpha$ denoting how the tension of the corresponding  brane scales with respect to the string coupling $g_S$. T-duality relates different potentials with the same value of $\alpha$. In particular, the RR potentials $C_p$ have $\alpha=-1$, while the potentials $D_{6+p,p}$ associated to the NS5-brane, the KK monopole and their T-duals have $\alpha=-2$ \cite{Bergshoeff:2011zk}. The potential $E_8$ associated to the S-dual of the D7-brane has $\alpha=-3$, and the other mixed-symmetry potentials with the same value of $\alpha$ are of the form $E_{8+n,2m,n}$ in IIB and $E_{8+n,2m+1,n}$ in IIA~\cite{Bergshoeff:2011ee}. In \cite{Lombardo:2016swq} it was shown that on the $T^6/[\mathbb{Z}_2 \times \mathbb{Z}_2 ]$ orientifold the IIB potential $E_8$ is mapped by mirror symmetry to specific components of the IIA potential $E_{9,3,1}$, and by adding also all the other components that can be consistently included in IIA, this maps in IIB to components of the potentials $E_{8,4}$, $E_{9,2,1}$ and $E_{10,4,2}$. Proceeding this way, one manages to obtain the full set of components of $\alpha=-3$ mixed-symmetry potentials corresponding to exotic branes of the IIB and IIA theory that can be consistently added in the $T^6/[\mathbb{Z}_2 \times \mathbb{Z}_2 ]$ orientifold model to cancel the tadpoles generated by the $\mathcal{P}$ fluxes.\footnote{All the components of the $\alpha=-3$ potentials corresponding to exotic branes in both the IIB and IIA  $T^6/[\mathbb{Z}_2 \times \mathbb{Z}_2 ]$ orientifold models are given in Table \ref{CEbranestable}.}

The aim of this paper is to complete the analysis carried out in \cite{Lombardo:2016swq} by including all possible non-geometric fluxes and exotic branes in the model. We start by considering the S-dual  of $\mathcal{P}^{a,b_1... b_4}$, which  following \cite{Aldazabal:2010ef} we denote as $\mathcal{Q}'^{a, b_1 ...b_4}$. 
In four dimensions this flux and those that are related to it by T-duality 
 have the same form of the NS fluxes, and we denote them as NS$'$ fluxes.
We manage to derive a universal T-duality rule that transforms the flux $\mathcal{Q}'^{a, b_1 ...b_4}$ to the other NS$'$ fluxes, which is given in eq. \eqref{TdualityruleH'fluxes} and actually applies to any dimension.\footnote{The highest dimension in which these fluxes appear is $D=7$.} Using this rule, we manage to determine the fluxes of the IIA theory that are mapped to the $\mathcal{Q}'^{a, b_1 ...b_4}$ fluxes of IIB. We then include all the additional IIA fluxes that are allowed, and we find that these are mapped in IIB to the components of the flux $\mathcal{H}'^{a_1 a_2 a_3, b_1 ... b_6}$. By S-duality, this latter flux is mapped to the flux that we denote as $\mathcal{F}'^{a_1 a_2 a_3, b_1... b_6}$. In four dimensions this flux and those that are related to it by T-duality 
are in the same representations of the RR fluxes, and we denote them as RR$'$ fluxes.
 We find a universal T-duality rule valid in any dimension,\footnote{Actually the highest dimension in which these fluxes appear is $D=4$.} which we give in eq. \eqref{TdualityruleF'fluxes}, that transforms the flux $\mathcal{F}'^{a_1 a_2 a_3, b_1... b_6}$ to additional RR$'$ fluxes. This rule allows us to derive the fluxes of the IIA orientifold model that are mapped to $\mathcal{F}'^{a_1 a_2 a_3, b_1... b_6}$  in IIB. The NS$'$ and RR$'$ fluxes, together with the RR, NS and $\mathcal{P}$ ones, give a total of 128 components, which are all the fluxes that can be included in the model.\footnote{All the NS$'$ and RR$'$  fluxes that can be turned on in the model, and how they are mapped by mirror symmetry, are given in Tables \ref{allH'fluxes} and \ref{allF'fluxes}.} The outcome of this analysis is that we manage to derive the most general expression for the superpotential for both the IIB and IIA theory, and in the IIB case this expression  coincides with the one derived in \cite{Aldazabal:2010ef}.

Having derived all the possible fluxes that can be turned on in the model, one can determine all the possible exotic branes that are sourced by these fluxes and thus can be included to cancel the tadpoles. We show that all the remaining branes that can be added have $\alpha=-5$ and $\alpha=-7$, and following \cite{Bergshoeff:2012jb} we denote the corresponding mixed-symmetry potentials as $G$ and $I$ respectively. This result is achieved by exploiting the universal T-duality rule derived in \cite{Lombardo:2016swq} which states as follows:
given  an $\alpha=-n$ brane associated to a mixed-symmetry potential such that the $a$ index occurs $p$ times (in $p$ different sets of antisymmetric indices),  this is mapped by T-duality along $a$ to the brane associated to the potential in which the $a$ index occurs $n-p$ times. Schematically, this can be written as
\begin{equation}
\alpha=-n \ : \qquad \quad \underbrace{a,a,...,a}_p \ \overset{{\rm T}_a}{\longleftrightarrow} \ \underbrace{a,a, ....,a }_{n-p} \quad . \label{allbranesallalphasruleintro}
\end{equation}
We proceed precisely as for the $\alpha=-3$ branes: the IIB potential $E_{10,4,2}$ is mapped by S-duality to $G_{10,4,2}$, and using eq. \eqref{allbranesallalphasruleintro} we map this by  mirror  symmetry to  components of the IIA potentials $G_{10,5,3,1}$, $G_{10,4,1}$ and $G_{10,6,5,2}$. We then add all the other components of these potentials that can consistently be included, and we map this back to IIB, resulting in the inclusion in this theory of the potentials $G_{10,5,4,1}$, $G_{10,6,2,2}$ and $G_{10,6,6,2}$. The latter potential is mapped by S-duality to $I_{10,6,6,2}$, which by mirror symmetry goes to components of the potential $I_{10,6,6,3}$ of IIA. Finally, the remaining component of this potential that can be added is mapped in IIB to $I_{10,6,6,6}$, which is a singlet under S-duality.\footnote{All the components of the $\alpha=-5$ and $\alpha=-7$ potentials corresponding to exotic branes in both IIB and IIA are given in Tables \ref{Gbranestable} and \ref{Ibranestable}.}

Having determined all the branes that can be included in the model, we can derive all the tadpole conditions for such branes. The quadratic terms in the fluxes that contribute to such tadpole conditions are schematically as follows:\footnote{In this formula and the next we schematically denote with $\mathcal{H}_{\rm NS}$ and $\mathcal{H}_{\rm NS'}$ all the NS and NS$'$ fluxes.}
\begin{align}
& \alpha=-1 \ \ {\rm potential} \ \ {C}:\  \ \   \mathcal{H}_{\rm NS} \cdot \mathcal{F}_{\rm RR} \nonumber \\
& \alpha=-3  \ \ {\rm potential} \ \  {E}:\ \ \   \mathcal{H}_{\rm NS} \cdot \mathcal{P} + \mathcal{F}_{\rm RR} \cdot \mathcal{H}_{\rm NS'}  \nonumber \\
& \alpha=-5 \ \ {\rm potential}\  \ {G}: \ \    \ \mathcal{H}_{\rm NS'} \cdot \mathcal{P} + \mathcal{F}_{\rm RR'} \cdot \mathcal{H}_{\rm NS} \nonumber \\
& \alpha=-7 \ \ {\rm potential} \ \  {I}:\ \ \ \   \mathcal{H}_{\rm NS'} \cdot \mathcal{F}_{\rm RR'}   \quad .\label{generalformoftadpolesintro}
\end{align}
Moreover, we determine all the branes of the maximal theory that are projected out in the orientifold and give non-trivial quadratic constraints for the fluxes. It turns out that these branes have even $\alpha$'s, and the resulting Bianchi identities have the schematic form
\begin{align}
& \alpha=-2 \ \ {\rm potential}:\  \ \   \mathcal{H}_{\rm NS} \cdot \mathcal{H}_{\rm NS} + \mathcal{P} \cdot \mathcal{F}_{\rm RR} =0\nonumber \\
& \alpha=-4  \ \ {\rm potential}:\ \ \   \mathcal{P} \cdot \mathcal{P} + \mathcal{H}_{\rm NS}\cdot \mathcal{H}_{\rm NS'} + \mathcal{F}_{\rm RR} \cdot \mathcal{F}_{\rm RR'}=0 \nonumber \\
& \alpha=-6 \ \ {\rm potential}: \ \    \ \mathcal{H}_{\rm NS'} \cdot \mathcal{H}_{\rm NS'}+ \mathcal{P} \cdot \mathcal{F}_{\rm RR'} =0   \quad . \label{Bianchibranesintro}
\end{align}
Again, by recursively applying mirror symmetry and S-duality, all such quadratic constraints are systematically  derived. 

The orientifold model we consider in this paper has a conjectured  $SL(2,\mathbb{Z})^7$ non-perturbative duality symmetry \cite{Aldazabal:2006up,Aldazabal:2008zza}, which can be understood by investigating the orbifold in more detail.
The 6-torus factorises as ${T^6 = \bigotimes_{i=1}^3 T_{(i)}^2}$, with the two $\mathbb{Z}_2$'s acting as $(-1,-1,1)$ and $(1,-1,-1)$ respectively on the coordinates ${(x^i , y^i )}$ of the three 2-tori, and there are seven untwisted moduli $S$, $T_i$ and $U_i$.
In the IIB/O3-orientifold model, the $S$ modulus is the axion-dilaton, the $T_i$ moduli are the  complex K\"{a}hler moduli which are given in terms of the  K\"{a}hler form and the RR 4-form, and the $U_i$ moduli are the complex structure moduli. In the low-energy supergravity theory in the absence of fluxes, each of these moduli parametrises the coset $SL(2,\mathbb{R})/SO(2)$, and the global symmetry $SL(2,\mathbb{R})^7$ is conjectured to be broken to  $SL(2,\mathbb{Z})^7$ in the full theory. The IIA/O6-orientifold arises from performing three T-dualities, one on each torus (following \cite{Aldazabal:2006up}, in this paper we  take these directions to be the three $x$ directions). The 128 fluxes that can be turned on in the model belong to the $({\bf 2,2,2,2,2,2,2})$ representation of $SL(2,\mathbb{R})^7$ \cite{Aldazabal:2006up}. In this paper we  determine the representations of all the branes that can be included. We  find that the branes belong to 16 irreducible representations, each made of three triplets and four singlets. More precisely, these 16 representations are all the possible representations made of three triplets and four singlets with the condition that in the IIB picture there is an even number of triplets with respect to the $SL(2,\mathbb{R})$'s associated to the $U_i$ moduli. Additionally, we find that the branes with even $\alpha$, that by being projected out give rise to the Bianchi identities, collect in 12 irreducible representations of $SL(2,\mathbb{R})^7$, which again are made of three triplets and four singlets.

The plan of the paper is as follows. In section 2 we determine the T-duality rules that transform the NS$'$ and RR$'$ fluxes, and we derive the expression for the superpotential with all possible fluxes turned on.  In section 3 we determine all the exotic branes that can be included in the model, and we derive all the tadpole conditions that result from turning on fluxes. We also derive all the Bianchi identities, which are the quadratic constraints that cannot be relaxed by the inclusion of branes. In section 4 we determine how all the exotic branes in the model transform with respect to the conjectured non-perturbative duality symmetry $SL(2,\mathbb{Z})^7$. Finally, section 5 contains our conclusions.

\begin{table}[h!]
\begin{center}
\scalebox{.85}{
\begin{tabular}{|c|c||c|c||c|}
\hline \multicolumn{2}{|c||}{IIB} & \multicolumn{2}{|c||}{IIA} & \# \\
 \cline{1-4} \rule[-1mm]{0mm}{2mm}  flux &  component & component & flux &  \\
\hline \hline \rule[-1mm]{0mm}{5mm} 
 $\mathcal{F}_3$ & $\mathcal{F}_{x^1 x^2 x^3}$ &   $\mathcal{F}$ & $\mathcal{F}_0$ & 1 \\
& $\mathcal{F}_{y^i x^j x^k}$ & $\mathcal{F}_{x^i y^i}$  & $\mathcal{F}_2$& 3 \\
& $ \mathcal{F}_{x^i y^j y^k}$ & $\mathcal{F}_{x^j y^j x^k y^k}$ & $\mathcal{F}_4$ & 3\\
\rule[-2.5mm]{0mm}{2mm}&  $\mathcal{F}_{y^1 y^2 y^3}$ & $\mathcal{F}_{x^1 y^1 x^2 y^2 x^3 y^3}$ & $\mathcal{F}_6$ & 1\\
 \hline
\hline \rule[3mm]{0mm}{1.5mm} 
 $\mathcal{H}_3$ & $\mathcal{H}_{x^1 x^2 x^3}$ &   ${ \mathcal{R}^{x^1 x^2 x^3}}$ & $\mathcal{R}$ & 1 \\
&  $\mathcal{H}_{y^i x^j x^k}$ & ${ -\mathcal{Q}_{y^i}^{x^j x^k}}$  & $\mathcal{Q}$ &  3 \\
& $ \mathcal{H}_{y^i y^j x^k}$ & $- f_{y^i y^j}^{x^k}$ & $f$ & 3\\
\rule[-2.5mm]{0mm}{2mm}
&  $\mathcal{H}_{y^1 y^2 y^3}$ & $\mathcal{H}_{y^1 y^2 y^3}$  & $\mathcal{H}_3$ &  1\\
 \hline \rule[3mm]{0mm}{1.5mm} 
$\mathcal{Q}$ & ${ \mathcal{Q}_{x^i}^{x^j x^k}}$ & $f_{x^j x^k}^{x^i}$  & $f$ & 3\\
& ${\mathcal{Q}_{x^i}^{x^j y^k}}$ & ${ \mathcal{Q}_{x^j}^{y^k x^i}}$ & $\mathcal{Q}$ & 3 \\
& ${\mathcal{Q}_{x^i}^{y^j x^k}}$ & ${\mathcal{Q}_{x^k}^{x^i y^j}}$ & $\mathcal{Q}$ & 3 \\

& ${ \mathcal{Q}_{y^i}^{x^j x^k}}$ & $-\mathcal{H}_{y^i x^j x^k}$ & $\mathcal{H}_3$ & 3\\
& ${\mathcal{Q}_{x^i}^{y^j y^k}}$ & ${ -\mathcal{R}^{ x^i y^j y^k}}$  & $\mathcal{R}$ & 3\\
& ${ \mathcal{Q}_{y^i}^{x^j y^k}}$ & $f_{y^i x^j}^{y^k}$ & $f$  & 3\\
& ${ \mathcal{Q}_{y^i}^{y^j x^k}}$ & $f_{x^k y^i}^{y^j}$ & $f$  & 3\\
\rule[-2.5mm]{0mm}{2mm}
&  ${\mathcal{Q}_{y^i}^{y^j y^k}}$ &  ${ \mathcal{Q}_{y^i}^{y^j y^k}}$ & $\mathcal{Q}$ & 3\\
\hline \hline \rule[3mm]{0mm}{1.5mm} $\mathcal{P}_1^2$& $\mathcal{P}_{x^i}^{x^j x^k}$ &$-\mathcal{P}^{x^i , x^i}$ & $\mathcal{P}^{1,1}$ & 3 \\
& $\mathcal{P}_{x^i}^{x^j y^k}$ & $\mathcal{P}^{x^i, x^i x^k y^k}$ & $\mathcal{P}^{1,3}$ & 3 \\
& $\mathcal{P}_{x^i}^{y^j x^k}$ & $\mathcal{P}^{x^i, x^i x^j y^j}$ & $\mathcal{P}^{1,3}$ & 3 \\
 & $\mathcal{P}_{y^i}^{x^j x^k}$ & $-\mathcal{P}_{y^i}^{x^i}$ & $\mathcal{P}_1^1$ & 3 \\
& $\mathcal{P}_{x^i}^{y^j y^k}$  & $\mathcal{P}^{x^i , x^i x^j x^k y^j y^k}$ & $\mathcal{P}^{1,5}$ & 3 \\ 
& $\mathcal{P}_{y^i}^{x^j y^k}$ & $\mathcal{P}_{y^i}^{x^i x^k y^k} $ & $\mathcal{P}_1^3$ & 3 \\
& $\mathcal{P}_{y^i}^{y^j x^k}$ & $\mathcal{P}_{y^i}^{x^i x^j y^j} $ & $\mathcal{P}_1^3$ & 3 \\
\rule[-2.5mm]{0mm}{2mm}
&  $\mathcal{P}_{y^i}^{y^j y^k}$     & $\mathcal{P}_{y^i}^{x^i x^j x^k y^j y^k}$ & $\mathcal{P}_1^5$ & 3 \\
\hline \rule[3mm]{0mm}{1.5mm}
$\mathcal{P}^{1,4}$ &  $\mathcal{P}^{x^i, y^i x^i x^j x^k }$     & $-\mathcal{P}_{x^i}^{y^i}$  & $\mathcal{P}_1^1$ & 3 \\
\rule[3mm]{0mm}{1.5mm}
&  $\mathcal{P}^{x^i, y^i x^i y^j x^k}$     & $\mathcal{P}_{x^i}^{y^i x^j y^j}$ & $\mathcal{P}_1^3$ & 3 \\
&  $\mathcal{P}^{x^i, y^i x^i x^j y^k}$     & $\mathcal{P}_{x^i}^{y^i x^k y^k}$ & $\mathcal{P}_1^3$ & 3 \\
&  $\mathcal{P}^{x^i, y^i x^i y^j y^k}$     & $\mathcal{P}_{x^i}^{y^i y^j y^k x^j x^k }$ & $\mathcal{P}_1^5$ &3 \\
 &  $\mathcal{P}^{y^i, x^i y^i x^j x^k }$     & $\mathcal{P}^{y^i,  y^i}$ & $\mathcal{P}^{1,1}$ & 3 \\
&  $\mathcal{P}^{y^i, x^i y^i y^j x^k}$     & $-\mathcal{P}^{y^i, y^i x^j y^j}$ & $\mathcal{P}^{1,3}$ &3 \\
&  $\mathcal{P}^{y^i, x^i y^i x^j y^k}$     & $-\mathcal{P}^{y^i, y^i x^k y^k}$ & $\mathcal{P}^{1,3}$ &3 \\
&  $\mathcal{P}^{y^i, x^i y^i y^j y^k }$     & $-\mathcal{P}^{y^i, y^i y^j y^k x^j x^k}$ & $\mathcal{P}^{1,5}$ &3 \\
\hline
\end{tabular}
}
\caption{\footnotesize All the RR, NS and $\mathcal{P}$ fluxes that can be turned on in the ${\cal N}=1$ orientifold model for both IIB and IIA. Here and in the rest of the paper, we choose the convention that to map the IIB to the mirror IIA theory we first perform a T-duality along $x^1$, then along $x^2$ and finally along $x^3$. The indices $i,j,k$ are always meant to be in cyclic order. In the last column we give the number of independent components for each flux. There are in total 8 RR fluxes, 32  NS fluxes and 48 $\mathcal{P}$ fluxes.}
\label{RRNSNSPfluxestable}
\end{center}
\end{table}

\section{NS$'$ $\&$ RR$'$ fluxes: T-duality and superpotential}
 
In this section we introduce the NS$'$ and RR$'$ fluxes, which are related by chains of T- and S-dualities to the RR and NS fluxes and the $\mathcal{P}$ fluxes recently considered in \cite{Lombardo:2016swq}. In particular, we will derive how these fluxes transform under T-duality and this will allow us to determine all the fluxes that can be turned on in the orientifold $T^6/[\mathbb{Z}_2 \times \mathbb{Z}_2 ]$ model for both the IIB and the IIA theory. This result will be used to derive the expression for   the $\mathcal{N}=1$ superpotential with all fluxes included for both theories. As it was the case for the analogous analysis carried out in \cite{Lombardo:2016swq} for the $\mathcal{P}$ fluxes, our IIB/O3 result fits with that found in \cite{Aldazabal:2010ef} on the basis of generalised geometry considerations and valid for any IIB/O3 orientifold with $SU(3)$ structure.
 
From the point of view of the four-dimensional effective action, fluxes give rise to gaugings, and in particular the fluxes of the ${\cal N}=1$ theory can be identified with specific components of the embedding tensor \cite{Nicolai:2000sc} of the maximal theory, which belongs to the ${\bf 912}$ of $E_{7(7)}$ \cite{deWit:2002vt}. 
 Under the branching $E_{7(7)}$ $\supset$ $SO(6,6)\times SL(2,\mathbb{R})$, where $SO(6,6)$  is the perturbative symmetry and  $SL(2,\mathbb{R})$  transforms non-linearly the complex scalar made of the four-dimensional dilaton and the axion dual to the NS 2-form, this representation decomposes as
 \begin{equation}
\textbf{912} = \textbf{(32,3)} \oplus  \textbf{(220,2)} \oplus  \textbf{(12,2)} \oplus  \textbf{(352,1)} \quad .
 \end{equation}
By further considering the embedding $SL(2,\mathbb{R})$ $\supset$ $\mathbb{R}^{+}$, where $\mathbb{R}^{+}$ is the dilaton weight, one finds that the various fluxes  belong to the following representations of the embedding tensor:
\begin{align} 
& {\rm RR \ fluxes}: \ \  \, \theta_\alpha \in {\bf  32_2} \nonumber \\
& {\rm NS \ fluxes}: \  \ \  \theta_{MNP} \in {\bf  220_1} \nonumber \\
& \mathcal{P} \ {\rm fluxes}: \ \ \ \ \, \theta_{M\dot{\alpha}} \in {\bf 352_0}  \label{SO66representationsfluxes} \\
& {\rm NS}' \ {\rm fluxes}: \ \ \theta'_{MNP} \in {\bf  220_{-1}} \nonumber \\
& {\rm RR}' \ {\rm fluxes}: \ \,\theta'_\alpha \in {\bf  32_{-2}} \quad . \nonumber 
\end{align} 
For the first three fluxes the form of the corresponding embedding tensor is the same in any dimension $D=10-d$, and the T-duality rules in eq. \eqref{TdualityruleRRfluxes}, \eqref{TdualityruleNSfluxes} and \eqref{TdualityrulePfluxes} can be easily understood as specific $O(d,d)$ transformations. In particular, observing that the $\mathcal{P}$ fluxes belong to a vector-spinor representation was crucial to derive the T-duality rules in eq. \eqref{TdualityrulePfluxes}, combining the transformation of the vector index  $M$, which splits in lower and upper $a$, with the transformation of the spinor index $\dot{\alpha}$,  which decomposes in the set of all even  or all odd antisymmetric indices $b$  \cite{Lombardo:2016swq}. 
We list in Table \ref{RRNSNSPfluxestable} all the RR, NS and $\mathcal{P}$ fluxes that can be turned on in the orientifold $T^6/[\mathbb{Z}_2 \times \mathbb{Z}_2 ]$ model for both the IIB and the IIA theory. In particular, the $\mathcal{P}$ fluxes of the IIA theory are determined by applying three T-dualities along the three $x$ directions  \cite{Lombardo:2016swq}.

The NS$'$ and RR$'$ fluxes collect in representations of $SO(d,d)$ that are not the same in any dimension.  In particular, the NS$'$ fluxes belong to the embedding tensor $\theta'_{M_1...M_{d-3}}$, and therefore they can be turned on in seven dimensions and below \cite{Bergshoeff:2012pm}. In four dimensions the embedding tensor  belongs to the representation with three antisymmetric indices, which decomposes in terms of the fluxes as 
\begin{equation}\label{H'}
\theta'_{M_1M_2M_3} \rightarrow \hspace{0.1 cm}  \mathcal{R}'^{b_1b_2b_3}  \hspace{0.2 cm} \mathcal{Q}'^{a_1,b_1b_2b_3b_4} \hspace{0.2 cm}  {f} '^{a_1a_2,b_1...b_5} \hspace{0.2 cm}\mathcal{H}'^{a_1a_2a_3,b_1...b_6} \quad .
\end{equation}
In this expression, the $a$ indices and the $b$ indices are separately completely antisymmetrised. Given that in general one can dualise $p$ upstairs indices of $SL(d,\mathbb{R})$ with $d-p$ downstairs ones, the reader can appreciate that the $SL(6,\mathbb{R})$ representations that occur in eq. \eqref{H'} are the same as those of the NS fluxes. On the other hand, only writing them with upstairs indices as in eq. \eqref{H'} one reproduces the correct embedding tensor in dimension higher than four. Indeed, in seven dimensions only $\mathcal{R}'$ can be turned on and it gives rise to a singlet $\theta'$. In six dimensions $\mathcal{R}'$ and $\mathcal{Q}'$ form the embedding tensor $\theta'_M$, and in five dimensions $\theta'_{M_1 M_2}$ is made out of $\mathcal{R}'$, $\mathcal{Q}'$ and $f'$. Finally, in three dimensions one has to also consider the flux $\tilde{\mathcal{H}}'^{a_1 ...a_4, b_1 ...b_7}$, which together with the fluxes in eq. \eqref{H'} gives rise to the embedding tensor  $\theta'_{M_1 ...M_4}$. The fact that the NS$'$ fluxes all have upstairs indices reveals their non-geometric nature, as also pointed out in \cite{Aldazabal:2010ef}. The index structure in eq. \eqref{H'} also follows naturally from the observation that these fluxes are dual to $\alpha=-4$ $(D-1)$-form potentials in $D$ dimensions, that originate from the ten-dimensional mixed-symmetry potentials $F_{9,3}$, $F_{9,4,1}$, $F_{9,5,2}$, $F_{9,6,3}$ and $F_{9,7,4}$.\footnote{Following \cite{Bergshoeff:2012ex}, we denote the potentials with $\alpha=-1,-2,-3...$ with the letters $C$, $D$, $E$ and so on.} 

Under a single T-duality the $\alpha=-4$ potentials transform as in eq. \eqref{allbranesallalphasruleintro} with $n=4$. Therefore, 
the duality between the NS$'$ fluxes and the $(D-1)$-form potentials with $\alpha=-4$ originating from the mixed-symmetry potentials above allows us to determine how a single T-duality transforms these fluxes in any dimension.
 The outcome of this analysis gives
\begin{equation}
\mathcal{R}'^{abc}  \overset{{\rm T}_d}{\longleftrightarrow} \mathcal{Q}'^{d,abcd}  \overset{{\rm T}_e}{\longleftrightarrow} f'^{ed,abcde}  \overset{{\rm T}_f}{\longleftrightarrow} \mathcal{H}'^{fed,abcdef} \label{TdualityruleH'fluxes}
\end{equation}
By contracting with an epsilon symbol of $SL(6,\mathbb{R})$, one can appreciate that in four dimensions these rules coincide with the ones that transform the NS fluxes. The flux $\mathcal{Q}'^{a, b_1 ...b_4}$ in IIB  is the S-dual of $\mathcal{P}^{a,b_1 ...b_4}$, which is connected by T-duality to $\mathcal{P}_a^{bc}$ and  by further S and T-dualities to the NS and RR  fluxes. As it is customary, in \cite{Lombardo:2016swq} all the components $f_{ab}^b$, $\mathcal{Q}_a^{ab}$ and $\mathcal{P}_a^{ab}$ (with indices not summed) were  put to zero. The T-duality rules in eq. \eqref{TdualityrulePfluxes} then map $\mathcal{P}_{a}^{bc}$ with $a$ different from $b$ and $c$ to  the components of $\mathcal{P}^{a,b_1 ...b_p}$ such that the $a$ index must coincide with one of the $b$ indices \cite{Lombardo:2016swq}.  We thus assume that the same occurs for the flux $\mathcal{Q}'^{a,b_1 ...b_4}$, and then the rules in eq. \eqref{TdualityruleH'fluxes} imply that for all the NS$'$ fluxes in eq. \eqref{H'} all the $a$ indices have to be parallel to some of the $b$ indices.

We now consider the RR$'$ fluxes, 
that in any dimension collect in the embedding tensor  $\theta'_{M_1 ...M_{d-6} \alpha}$ of $SO(d,d)$. This can actually only be defined in four dimensions and below, and in particular in four dimensions we decompose it in terms of fluxes as\,\footnote{\label{footnote} In eq. \eqref{decompositionofRR'fluxes} we only write down the fluxes that are relevant in four dimensions. In $D=3$ one also has the fluxes $\mathcal{F}'^{a,b_1 .. b_p ,c_1 ...c_7}$ with $p$ odd in IIB and even in IIA.} 
\begin{equation}
\theta_\alpha'  \rightarrow \left\{ \begin{array}{ll} \mathcal{F}'^{a_1,b_1...b_6} \ \ \mathcal{F}'^{a_1a_2a_3,b_1...b_6} \ \    \mathcal{F}'^{a_1...a_5,b_1...b_6} \ \  \ \ \  \ \ \ \ \ \ \ \ \ \ \ ({\rm IIB})
        \\
        \\
         \mathcal{F}'^{b_1...b_6} \ \  \mathcal{F}'^{a_1a_2,b_1...b_6} \ \  \mathcal{F}'^{a_1...a_4,b_1...b_6}  \ \ \mathcal{F}'^{a_1...a_6,b_1...b_6} \ \ \ ({\rm IIA})\end{array} \right. \quad ,  \quad  \label{decompositionofRR'fluxes}
\end{equation}
and using the $SL(6,\mathbb{R})$ epsilon symbol one can see that the representations that occur are the same as those of the RR fluxes. Again, all these fluxes are non-geometric and the index structure is motivated by the duality with the $(D-1)$-form potentials $G$ and by requiring that they give rise to the correct embedding tensor also in dimension lower than four, provided that one also includes the contribution of the additional fluxes mentioned in footnote \ref{footnote}. 

The T-duality rules for the RR$'$ fluxes in four dimensions are
\begin{equation}
\mathcal{F}'^{a_1...a_6} \overset{{\rm T}_{a_1}}{\longleftrightarrow} \mathcal{F}'^{a_1,a_1...a_6} \overset{{\rm T}_{a_2}}{\longleftrightarrow} \mathcal{F}'^{a_1a_2,a_1...a_6} \hspace{0.1 cm} ... \hspace{0.1 cm} \overset{{\rm T}_{a_6}}{\longleftrightarrow} \mathcal{F}'^{a_1...a_6,a_1...a_6} \quad .  \label{TdualityruleF'fluxes}
\end{equation}
In three dimensions, one must add the additional T-duality rule 
\begin{equation}
\mathcal{F}'^{a_1 ...a_p, b_1...b_6}  \overset{{\rm T}_{c}}{\longleftrightarrow} \mathcal{F}'^{c,a_1 ... a_p c,b_1...b_6 c}
\end{equation}
for the transformation along a direction that is not present in the flux. As for the case of the NS$'$ fluxes, we only consider components such that in eq. \eqref{decompositionofRR'fluxes}  all the $a$ indices are parallel to some of the $b$ indices, which is consistent with the T-duality rules in eq. \eqref{TdualityruleF'fluxes}.

We can now apply the T-duality rules in eqs. \eqref{TdualityruleH'fluxes} and \eqref{TdualityruleF'fluxes} 
to determine all the NS$'$ and RR$'$ fluxes that can be included in the four-dimensional $T^6/[\mathbb{Z}_2 \times \mathbb{Z}_2 ]$ orientifold model. By S-duality, the  components of the $\mathcal{P}^{1,4}$ flux of the IIB theory that are listed in the last eight rows of Table \ref{RRNSNSPfluxestable} are mapped to the same components of the $\mathcal{Q}'^{1,4}$ flux. By applying three T-dualities along the $x$ directions using eq. \eqref{TdualityruleH'fluxes}, one gets the IIA fluxes that are listed on the right-hand side of the first eight rows of Table \ref{allH'fluxes}. In the IIA theory, the orientifold projection selects  components of $\mathcal{R}'$ and $f'$ with an odd number of $y$'s and components of $\mathcal{Q}'$ and $\mathcal{H}'$ with an even number of $y$'s. By adding all the other components that satisfy this criterion and are therefore compatible with the orientifold, one also includes the last four IIA fluxes in Table \ref{allH'fluxes}, which are mapped in IIB to all the allowed components of the $\mathcal{H}'$ flux.

  \begin{table}[t!]
\begin{center}
\begin{tabular}{|c|c||c|c||c|}
\hline \multicolumn{2}{|c||}{IIB} & \multicolumn{2}{|c||}{IIA} & \# \\
 \cline{1-4} \rule[-1mm]{0mm}{2mm}  flux &  component & component & flux &  \\
\hline 
\hline \rule[-1mm]{0mm}{6mm} $\mathcal{Q}'^{1,4}$  & $\mathcal{Q}'^{x^i, y^i x^i x^j x^k }$     & $\mathcal{R}'^{y^i x^j x^k }$ &$\mathcal{R}'^3$ & 3  \\
 & $\mathcal{Q}'^{x^i, y^i x^i x^j y^k}$     & $\mathcal{Q}'^{x^k, y^k y^i x^j x^k }$ &$\mathcal{Q}'^{1,4}$ & 3  \\
 & $\mathcal{Q}'^{x^i, y^i x^i y^j x^k}$     & $\mathcal{Q}'^{x^j, y^j y^i x^j x^k }$ &$\mathcal{Q}'^{1,4}$ & 3  \\
   &  $\mathcal{Q}'^{x^i,y^i x^i  y^j y^k}$     & $- f'^{x^j x^k, y^i y^j y^k x^j x^k}$ & $f'^{2,5}$ & 3 \\
   & $\mathcal{Q}'^{y^i, x^i y^i x^j x^k }$     & $\mathcal{Q}'^{y^i, x^i y^i x^j x^k }$   & $\mathcal{Q}'^{1,4}$ & 3\\

  &  $\mathcal{Q}'^{y^i, x^i y^i  x^j y^k}$     & $f'^{x^k y^i, x^i y^i x^j y^k x^k}$  & $f'^{2,5}$ & 3 \\

  &  $\mathcal{Q}'^{y^i, x^i  y^i y^j x^k}$     & $f'^{x^j y^i, x^i y^i y^j x^k x^j}$  & $f'^{2,5}$ & 3 \\

 \rule[-1mm]{0mm}{2mm}  & $\mathcal{Q}'^{y^i, x^i y^i y^j y^k }$     & $- \mathcal{H}'^{y^i x^j x^k , x^1 x^2 x^3 y^1 y^2 y^3}$  & $\mathcal{H}'^{3,6}$ & 3  \\
 \hline \rule[-1mm]{0mm}{6mm} 
$\mathcal{H}'^{3,6}$ & $\mathcal{H}'^{x^1 x^2 x^3,x^1 x^2 x^3 y^1 y^2 y^3 }$ & $\mathcal{R}'^{y^1 y^2 y^3}$ & $\mathcal{R}'^{3}$ & 1  \\
  & $\mathcal{H}'^{y^i x^j x^k ,x^1 x^2 x^3 y^1 y^2 y^3 }$ & $- \mathcal{Q}'^{y^i,x^i y^i y^j y^k}$  & $\mathcal{Q}'^{1,4}$ & 3  \\
  & $\mathcal{H}'^{x^i y^j y^k,x^1 x^2 x^3 y^1 y^2 y^3}$ &  $- f'^{y^j y^k, y^i y^j y^k x^j x^k }$   & $f'^{2,5}$ & 3   \\
 & $\mathcal{H}'^{y^1 y^2 y^3,x^1 x^2 x^3 y^1 y^2 y^3}$ & $\mathcal{H}'^{y^1 y^2 y^3 , x^1 x^2 x^3  y^1 y^2 y^3}$  & $\mathcal{H}'^{3,6}$ & 1 \\
\hline
\end{tabular}
\caption{\footnotesize Table containing all the NS$'$ fluxes that can be turned on in the ${\cal N}=1$ orientifold model for both IIB and IIA.  In the last column we give the number of independent components for each flux, whose total number is 32.}
\label{allH'fluxes}
\end{center}
\end{table}

One can similarly determine all the allowed RR$'$ fluxes in the orientifold. By S-duality, the components of the $\mathcal{H}'$ flux in the IIB theory are mapped to the same components of the $\mathcal{F}'^{3,6}$ flux. By applying three T-dualities along the $x$ directions using eq. \eqref{TdualityruleF'fluxes}, one derives the components of the RR$'$ fluxes that are allowed in the mirror IIA orientifold. The result is given  in Table \ref{allF'fluxes}. From the table one can deduce that the orientifold selects the components of $\mathcal{F}'^6$ and $\mathcal{F}'^{4,6}$ with an odd number of $y$'s and the components of $\mathcal{F}'^{2,6}$ and $\mathcal{F}'^{6,6}$ with an even number of $y$'s. To summarise, Tables \ref{RRNSNSPfluxestable}, \ref{allH'fluxes} and \ref{allF'fluxes} give all the possible fluxes that can be included in the model, for both the IIB and the IIA theory. 
The aim of the remaining of this section is to write down an expression for the  superpotential for both the IIB and the IIA theory with all these fluxes turned on.

\begin{table}[t!]
\begin{center}
\begin{tabular}{|c|c||c|c||c|}
\hline \multicolumn{2}{|c||}{IIB} & \multicolumn{2}{|c||}{IIA} & \# \\
 \cline{1-4} \rule[-1mm]{0mm}{2mm}  flux &  component & component & flux &  \\
\hline 
\hline \rule[-1mm]{0mm}{6mm}
$\mathcal{F}'^{3,6}$  & $\mathcal{F}'^{x^1 x^2 x^3 , x^1 x^2 x^3 y^1 y^2 y^3}$ & $- \mathcal{F}'^{x^1 x^2 x^3 y^1 y^2 y^3 }$  & $\mathcal{F}'^6$& 1 \\ 
  & $\mathcal{F}'^{y^i x^jx^k,x^1 x^2 x^3 y^1 y^2 y^3 }$ & $\mathcal{F}'^{ x^i y^i, x^1 x^2 x^3 y^1 y^2 y^3}$ & $\mathcal{F}'^{2,6}$ &  3 \\
 & $\mathcal{F}'^{x^i y^j y^k , x^1 x^2 x^3 y^1 y^2 y^3 }$ & $-\mathcal{F}'^{x^j y^j x^k y^k , x^1 x^2 x^3 y^1 y^2 y^3}$  & $\mathcal{F}'^{4,6}$ & 3   \\
   & $\mathcal{F}'^{y^1 y^2 y^3,x^1 x^2 x^3 y^1 y^2 y^3}$ &  $- \mathcal{F}'^{x^1 x^2 x^3 y^1 y^2 y^3 , x^1 x^2 x^3 y^1 y^2 y^3}$ & $\mathcal{F}'^{6,6}$ & 1  \\
\hline
\end{tabular}
\caption{\footnotesize Table containing all the RR$'$ fluxes that can be turned on in the ${\cal N}=1$ orientifold model for both IIB and IIA. In the last column we give the number of independent components for each flux, whose total number is 8. }
\label{allF'fluxes}
\end{center}
\end{table}
 
In order to derive the superpotential in the orientifold model, we give  the explicit expression of the holomorphic 3-form and the K\"{a}hler form as functions of the moduli following the conventions of \cite{Aldazabal:2006up}. In IIB the non-vanishing components of the 3-form $\Omega$ are 
\begin{equation}
\Omega_{x^1 x^2 x^3 } = 1 \qquad \Omega_{y^i x^j x^k } = i \, U_i \qquad \Omega_{x^i y^j y^k} = - U_j U_k \qquad \Omega_{y^1 y^2 y^3} = -i\, U_1 U_2 U_3 \quad , \label{modulidep1IIB}
\end{equation}
while the non-vanishing components of the complexified K\"{a}hler 4-form are
\begin{equation}
(\mathcal{J}_{{\rm c}})_{x^j y^j x^k y^k } = i\, T_i \quad .\label{modulidep2IIB}
\end{equation}
In IIA, the only non-vanishing components of the complexified holomorphic 3-form and  the complexified K\"{a}hler 2-form are
\begin{equation}
(\Omega_{{\rm c}})_{  x^1 x^2 x^3 } = i\, S   \qquad (\Omega_{{\rm c}})_{ x^i y^j y^k} = - i \, U_i \qquad 
(J_{{\rm c}})_{ x^i y^i} = -i\, T_i \quad .\label{modulidepIIA}
\end{equation}
In all these expressions, as everywhere else in this section, the indices $i$, $j$ and $k$ are meant to be in cyclic order. The two theories are mapped into each other by 
performing a T-duality transformation along $x^1$ followed by one along $x^2 $ and one along $x^3$, corresponding to localising the O6-plane in the $y$ directions. The resulting expressions for the superpotential in the two theories are identified via the exchange of the $T$ and $U$ moduli.

The superpotential in the presence of non-geometric  NS fluxes was originally derived in \cite{Shelton:2005cf} by applying the NS T-duality rules in eq. \eqref{TdualityruleNSfluxes} to the Gukov-Vafa-Witten  superpotential in IIB \cite{Gukov:1999ya} and to the one derived in \cite{Derendinger:2004jn,Villadoro:2005cu} in IIA, where all the possible geometric fluxes were included. The expression for the IIB superpotential with all RR and NS fluxes included is 
\begin{equation}
W_{{\rm IIB/O3},{\rm RR,NS}} = \int [ \mathcal{F}_3 - i S \mathcal{H}_3    +  \mathcal{Q}  \cdot \mathcal{J}_{\rm c} ] \wedge \Omega \quad , \label{superpotIIBRRNSfluxes}
\end{equation}
where 
\begin{equation}
(\mathcal{Q} \cdot \mathcal{J}_{\rm c})_{a_1a_2a_3}=\tfrac{3}{2} \cdot \mathcal{Q}^{b_1b_2}_{[a_1}(\mathcal{J}_{\rm c})_{a_2a_3]b_1b_2} \quad .\label{contractionNSIIB}
\end{equation}
In IIA, the terms containing the NS fluxes were written in \cite{Aldazabal:2006up} in the form $\Omega_{\rm c} \wedge {\rm flux} \cdot J_{\rm c}^n$, with $n=0,...,3$, with the upstairs indices of the fluxes contracting indices of $J_{\rm c}^n$. Using Fierz identities, one can show that these terms  can also be rewritten in the form $e^{J_{\rm c}} \wedge \mathcal{H}_{\rm NS} \cdot \Omega_{\rm c}$, with the upstairs indices of the fluxes contracting indices of $\Omega_{\rm c}$ instead of indices of ${J_{\rm c}}$. We therefore write the IIA superpotential with all RR and NS fluxes turned on as
\begin{align}
W_{{\rm IIA/O6},{\rm RR,NS}} & = \int e^{J_{\rm c}} \wedge [ \mathcal{F}_{\rm RR} - \mathcal{H}_{\rm NS} \cdot \Omega_{\rm c} ]\nonumber \\
& = \int [ \mathcal{F}_6 - \mathcal{H}_3 \wedge \Omega_{\rm c} + {J_{\rm c}} \wedge ( \mathcal{F}_4  - f \cdot \Omega_{\rm c}  ) 
+ \tfrac{1}{2} {J_{\rm c}} \wedge {J_{\rm c}} \wedge  ( \mathcal{F}_2 -  \mathcal{Q} \cdot \Omega_{\rm c}  ) \nonumber \\
 & + \tfrac{1}{6} {J_{\rm c}}\wedge {J_{\rm c}}\wedge {J_{\rm c}} \, ( \mathcal{F}_0 - \mathcal{R} \cdot \Omega_{\rm c})] \quad ,
\label{superpotIIARRNSfluxes}
\end{align}
where the contractions are defined as
\begin{align}
& (f \cdot \Omega_{\rm c})_{a_1...a_4}= 6 \cdot f^{b}_{[a_1a_2}(\Omega_{\rm c})_{a_3a_4]b}\nonumber \\
& (\mathcal{Q} \cdot \Omega_{\rm c})_{a_1a_2}=\mathcal{Q}^{b_1b_2}_{[a_1}(\Omega_{\rm c})_{a_2]b_1b_2}\label{contractionsNSIIA}\\
&  \mathcal{R} \cdot \Omega_{\rm c} =\tfrac{1}{6} \cdot \mathcal{R}^{b_1b_2b_3}(\Omega_{\rm c})_{b_1b_2b_3} \quad . \nonumber 
\end{align}
Identifying the fluxes in the two theories as in Table \ref{RRNSNSPfluxestable},  eqs. \eqref{superpotIIBRRNSfluxes} and \eqref{superpotIIARRNSfluxes}  are mapped into each other by swapping the moduli $T$ and $U$. In particular, using eqs. \eqref{modulidep1IIB}, \eqref{modulidep2IIB} and \eqref{modulidepIIA} one can see that  the superpotential is of the form ${P}_1 (U) + S { P}_2 (U) + T { P}_3 (U)$ in IIB and ${P}_1 (T) + S { P}_2 (T) + U { P}_3 (T)$ in IIA, where the ${ P}$'s are all cubic polynomials.

The contribution of the  $\mathcal{P}_1^2$ flux in IIB was  considered in \cite{Aldazabal:2006up} using S-duality. This induces a term $ST { P}_4 (U) $ in the superpotential, and the mirror $SU {P}_4 (T)$ expression in IIA  arises from a term that can be schematically written as $(J_{\rm c})^n \cdot \mathcal{P} \cdot \Omega_{\rm c}^2$, with $n= 0,...,3$. 
In \cite{Lombardo:2016swq} the T-duality rules in eq. \eqref{TdualityrulePfluxes} were used to derive explicitly such expression, also giving rise to a term $U^2 {P}_5(T)$, which is mapped back to IIB to a term $T^2 { P}_5 (U)$ originating from the flux $\mathcal{P}^{1,4}$. As a result, the full expression for the part of the IIB superpotential containing the $\mathcal{P}$ fluxes is 
\begin{equation}
W_{{\rm IIB/O3},\mathcal{P}} = \int [ - i S  \mathcal{P}_1^2  \cdot \mathcal{J}_{\rm c} + \mathcal{P}^{1,4} \cdot \mathcal{J}_{\rm c}^2 ]\wedge \Omega \quad ,
\label{Wballpfluxes} 
\end{equation}
where the contractions are 
\begin{align}
& (\mathcal{P}_1^2 \cdot \mathcal{J}_{\rm c})_{a_1a_2a_3}=\tfrac{3}{2} \cdot \mathcal{P}^{b_1b_2}_{[a_1}(\mathcal{J}_{\rm c})_{a_2a_3]b_1b_2} \nonumber \\
& (\mathcal{P}^{1,4} \cdot \mathcal{J}_{\rm c}^2)_{a_1a_2a_3}= \tfrac{1}{4} \cdot \mathcal{P}^{c,b_1...b_4}(\mathcal{J}_{\rm c})_{[a_1a_2|c b_1|}(\mathcal{J}_{\rm c})_{a_3]b_2b_3b_4} \quad . \label{contractionsPIIB}
\end{align}
Precisely as done above for the terms containing the NS fluxes, the IIA mirror of eq. \eqref{Wballpfluxes} given in \cite{Lombardo:2016swq} can be rewritten using Fierz identities in the form $e^{J_{\rm c}} \wedge \mathcal{P} \cdot \Omega_{\rm c}^2$. Moreover, one can show that, modulo Fierz rearrangements, the resulting expression is unique. We therefore write down the $\mathcal{P}$-flux part of the IIA superpotential as 
\begin{align}
W_{{\rm IIA/O6},\mathcal{P}} & = \int e^{J_{\rm c}} \wedge \mathcal{P} \cdot \Omega_{\rm c}^2=\int [ \mathcal{P}^1_1 \cdot \Omega_{\rm c}^2+ J_{\rm c} \wedge (\mathcal{P}^{1,1}+\mathcal{P}_1^3) \cdot \Omega_{\rm c}^2 \nonumber \\
& + \tfrac{1}{2} J_{\rm c} \wedge J_{\rm c} \wedge (\mathcal{P}_1^5+\mathcal{P}^{1,3}) \cdot \Omega_{\rm c}^2 + \tfrac{1}{6} J_{\rm c} \wedge J_{\rm c} \wedge J_{\rm c}  \ \mathcal{P}^{1,5} \cdot \Omega_{\rm c}^2 \, ]\quad ,
\label{Waallpfluxes}
\end{align}
where the contractions are defined as 
\begin{align}
& (\mathcal{P}^1_1 \cdot \Omega_{\rm c}^2)_{a_1...a_6}=60 \cdot \tfrac{1}{2} \cdot \mathcal{P}^b_{[a_1}(\Omega_{\rm c})_{a_2a_3|b|}(\Omega_{\rm c})_{a_4a_5a_6]}\nonumber \\
& 
(\mathcal{P}^{1,1} \cdot \Omega_{\rm c}^2)_{a_1a_2a_3a_4}= - 6 \cdot \tfrac{1}{2} \cdot  \mathcal{P}^{c,b}(\Omega_{\rm c})_{c[a_1a_2}(\Omega_{\rm c})_{a_3a_4]b} \nonumber \\
& (\mathcal{P}_1^3 \cdot \Omega_{\rm c}^2)_{a_1a_2a_3a_4}= - 6 \cdot \tfrac{1}{2} \cdot \mathcal{P}_{[a_1}^{b_1b_2b_3}(\Omega_{\rm c})_{a_2a_3|b_1|}(\Omega_{\rm c})_{a_4]b_2b_3}\nonumber \\
&(\mathcal{P}^{1,3} \cdot \Omega_{\rm c}^2)_{a_1a_2}=\tfrac{1}{2} \cdot \mathcal{P}^{c,b_1b_2b_3}(\Omega_{\rm c})_{b_1b_2[a_1}(\Omega_{\rm c})_{a_2]c b_3} \nonumber \\
&(\mathcal{P}_1^5 \cdot \Omega_{\rm c}^2)_{a_1a_2}= \tfrac{1}{6} \cdot \tfrac{1}{2} \cdot \mathcal{P}_{[a_1}^{b_1...b_5}(\Omega_{\rm c})_{a_2]b_1b_2}(\Omega_{\rm c})_{b_3b_4b_5} \nonumber \\
& \mathcal{P}^{1,5} \cdot \Omega_{\rm c}^2= -\tfrac{1}{12} \cdot \tfrac{1}{2} \cdot \mathcal{P}^{c,b_1...b_5}(\Omega_{\rm c})_{b_1b_2b_3}(\Omega_{\rm c})_{b_4b_5c} \quad . \label{contractionsPIIA}
\end{align}
The reader can check that using eqs. \eqref{modulidep1IIB}, \eqref{modulidep2IIB} and \eqref{modulidepIIA} and the relations among the $\mathcal{P}$ fluxes given in Table \ref{RRNSNSPfluxestable} the IIA expression is the mirror of the IIB one.

We want to generalise the analysis  to include all the allowed fluxes. We first consider the NS$'$ fluxes. In IIB,  $\mathcal{Q}'^{1,4}$ is the S-dual of $\mathcal{P}^{1,4}$, and one can immediately write down its contribution to the superpotential. One then maps this to IIA using the relations among the fluxes given in Table \ref{allH'fluxes}. The IIA expression for the superpotential also contains terms that are mapped back in IIB to $\mathcal{H}'^{3,6}$ terms. The resulting IIB superpotential is 
 \begin{equation}
W_{{\rm IIB/O3},{\rm NS'}}  =\int [ -iS \mathcal{Q}'^{1,4}\cdot \mathcal{J}_{\rm c}^2 + \mathcal{H}'^{3,6} \cdot \mathcal{J}^3_{\rm c}]\wedge \Omega \quad ,
\end{equation}
where the contractions are defined as
\begin{align}
& (\mathcal{Q}'^{1,4} \cdot \mathcal{J}_{\rm c}^2)_{a_1a_2a_3}= \tfrac{1}{4} \cdot \mathcal{Q}'^{c,b_1...b_4}(\mathcal{J}_{\rm c})_{[a_1a_2|c b_1|}(\mathcal{J}_{\rm c})_{a_3]b_2b_3b_4} \nonumber \\
& (\mathcal{H}'^{3,6} \cdot \mathcal{J}_{\rm c}^3)_{a_1a_2a_3}=\tfrac{1}{32} \cdot \tfrac{1}{6} \cdot \mathcal{H}'^{c_1c_2c_3,b_1...b_6}(\mathcal{J}_{\rm c})_{[a_1a_2|c_1c_2|}(\mathcal{J}_{\rm c})_{a_3]c_3b_1b_2}(\mathcal{J}_{\rm c})_{b_3b_4b_5b_6} \label{contractionsNS'IIB}
\end{align}
In IIA, we again find that modulo Fierz rearrangements there is a unique expression that one can write in the form  $e^{J_{\rm c}} \wedge \mathcal{H}_{\rm NS'} \cdot \Omega_{\rm c}^3$, which is 
\begin{align}
W_{{\rm IIA/O6},{\rm NS'}} & =\int [-e^{J_{\rm c}} \wedge \mathcal{H}_{\rm NS'} \cdot \Omega_{\rm c}^3 ]= \int [ - \mathcal{R}'^3 \cdot \Omega_{\rm c}^3 - J_{\rm c} \wedge \mathcal{Q}'^{1,4} \cdot \Omega_{\rm c}^3\nonumber \\
& - \tfrac{1}{2} J_{\rm c} \wedge J_{\rm c} \wedge f'^{2,5} \cdot \Omega_{\rm c}^3 -\tfrac{1}{6} J_{\rm c} \wedge J_{\rm c} \wedge J_{\rm c}\ \mathcal{H}'^{3,6} \cdot \Omega_{\rm c}^3 \, ]\quad ,
\end{align}
where the contractions are defined as
\begin{align}
& (\mathcal{R}'^3 \cdot \Omega_{\rm c}^3)_{a_1...a_6}=  30 \cdot \tfrac{1}{6} \cdot \mathcal{R}'^{b_1b_2b_3}(\Omega_{\rm c})_{[a_1a_2a_3}(\Omega_{\rm c})_{a_4a_5|b_1|}(\Omega_{\rm c})_{a_6]b_2b_3}\nonumber \\
& (\mathcal{Q}'^{1,4} \cdot \Omega_{\rm c}^3)_{a_1...a_4}=\tfrac{1}{2} \cdot \mathcal{Q}'^{c,b_1...b_4}(\Omega_{\rm c})_{[a_1a_2|c}(\Omega_{\rm c})_{a_3a_4]b_1}(\Omega_{\rm c})_{b_2b_3b_4} \nonumber \\
& (f'^{2,5} \cdot \Omega_{\rm c}^3)_{a_1a_2}=-\tfrac{1}{12} \cdot \tfrac{1}{2} \cdot  f'^{c_1c_2,b_1...b_5}(\Omega_{\rm c})_{[a_1|c_1c_2|}(\Omega_{\rm c})_{a_2]b_1b_2}(\Omega_{\rm c})_{b_3b_4b_5} \nonumber \\
& \mathcal{H}'^{3,6} \cdot \Omega_{\rm c}^3 = -\tfrac{1}{24} \cdot \tfrac{1}{6} \mathcal{H}'^{c_1c_2c_3,b_1...b_6}(\Omega_{\rm c})_{c_1c_2b_1}(\Omega_{\rm c})_{c_3b_2b_3}(\Omega_{\rm c})_{b_4b_5b_6} \quad . \label{contractionsNS'IIA}
\end{align}
Again, it can be checked that  using eqs. \eqref{modulidep1IIB}, \eqref{modulidep2IIB} and \eqref{modulidepIIA} and the relations among the NS$'$ fluxes given in Table \ref{allH'fluxes} the IIA and IIB expressions are related by mirror symmetry, with the IIB polynomial which has the form $ST^2 {P}_6 (U) + T^3 {P}_7 (U)$.

We finally consider the RR$'$ fluxes. In IIB, the only contribution comes from $\mathcal{F}'^{3,6}$, which is the S-dual of $\mathcal{H}'^{3,6}$. This results in the contribution 
 \begin{equation}
W_{{\rm IIB/O3},{\rm RR'}}  =\int [ -iS  \mathcal{F}'^{3,6} \cdot \mathcal{J}^3_{\rm c}]\wedge \Omega 
\end{equation}
to the superpotential, where  the contraction is defined as 
\begin{equation}
(\mathcal{F}'^{3,6} \cdot \mathcal{J}_{\rm c}^3)_{a_1a_2a_3}=\tfrac{1}{32} \cdot \tfrac{1}{6} \cdot \mathcal{F}'^{c_1c_2c_3,b_1...b_6}(\mathcal{J}_{\rm c})_{[a_1a_2|c_1c_2|}(\mathcal{J}_{\rm c})_{a_3]c_3b_1b_2}(\mathcal{J}_{\rm c})_{b_3b_4b_5b_6}\quad .\label{contractionRR'IIB}
\end{equation}
Using Table \ref{allF'fluxes}, this is mapped in IIA to the expression
\begin{align}
W_{{\rm IIA/O6},{\rm RR}'} &= \int  e^{J_{\rm c}} \wedge \mathcal{F}_{\rm RR'} \cdot \Omega_{\rm c}^4 =
\int [ \mathcal{F}'^{6} \cdot \Omega_{\rm c}^4 + J_{\rm c} \wedge \mathcal{F}'^{2,6} \cdot \Omega_{\rm c}^4\nonumber \\
& + \tfrac{1}{2} J_{\rm c} \wedge J_{\rm c} \wedge \mathcal{F}'^{4,6} \cdot \Omega_{\rm c}^4+ \tfrac{1}{6} J_{\rm c} \wedge J_{\rm c} \wedge J_{\rm c}  \, \mathcal{F}'^{6,6} \cdot \Omega_{\rm c}^4 \, ] \quad ,
\end{align}
where the contractions are defined as
\begin{align}
& (\mathcal{F}'^6 \cdot \Omega_{\rm c}^4)_{a_1...a_6}=5 \cdot \tfrac{1}{24} \cdot \mathcal{F}'^{b_1...b_6}(\Omega_{\rm c})_{[a_1a_2a_3}(\Omega_{\rm c})_{a_4a_5|b_1|}(\Omega_{\rm c})_{a_6]b_2b_3}(\Omega_{\rm c})_{b_4b_5b_6}\nonumber \\
& (\mathcal{F}'^{2,6} \cdot \Omega_{\rm c}^4)_{a_1...a_4}=\tfrac{1}{2} \cdot \tfrac{1}{8}\cdot  \mathcal{F}'^{c_1c_2,b_1...b_6}(\Omega_{\rm c})_{[a_1a_2|c_1|}(\Omega_{\rm c})_{a_3a_4]b_1}(\Omega_{\rm c})_{c_2b_2b_3}(\Omega_{\rm c})_{b_4b_5b_6}\nonumber \\
& (\mathcal{F}'^{4,6} \cdot \Omega_{\rm c}^4)_{a_1a_2}=\tfrac{1}{12} \cdot \tfrac{1}{16} \cdot \mathcal{F}'^{c_1...c_4,b_1...b_6}(\Omega_{\rm c})_{[a_1|c_1c_2|}(\Omega_{\rm c})_{a_2]c_3b_1}(\Omega_{\rm c})_{c_4b_2b_3}(\Omega_{\rm c})_{b_4b_5b_6}\nonumber \\
& \mathcal{F}'^{6,6} \cdot \Omega_{\rm c}^4=\tfrac{1}{144} \cdot \tfrac{1}{24} \cdot \mathcal{F}'^{c_1...c_6,b_1...b_6}(\Omega_{\rm c})_{c_1c_2c_3}(\Omega_{\rm c})_{c_4c_5b_1}(\Omega_{\rm c})_{c_6b_2b_3}(\Omega_{\rm c})_{b_4b_5b_6} \quad . \label{contractionsRR'IIA}
\end{align}
Again, mirror symmetry relates the IIB and IIA expressions, and in particular in IIB one gets a contribution to the superpotential of the form $S T^3 {P}_8 (U)$. 

It is instructive to summarise our results writing down the superpotential with all fluxes included for both theories. In IIB one gets
\begin{align}
W_{{\rm IIB/O3}} & = \int [\mathcal{F}_3 - i S \mathcal{H}_3  +  ( \mathcal{Q}  - i S  \mathcal{P}_1^2 ) \cdot \mathcal{J}_{\rm c} + ( \mathcal{P}^{1,4}  -iS \mathcal{Q}'^{1,4} )\cdot \mathcal{J}_{\rm c}^2 \nonumber \\
&+ ( \mathcal{H}'^{3,6}-iS  \mathcal{F}'^{3,6}) \cdot \mathcal{J}^3_{\rm c} ] \wedge \Omega \quad , \label{superpotIIBfull}
\end{align}
where the contractions are given in eqs. \eqref{contractionNSIIB}, \eqref{contractionsPIIB}, \eqref{contractionsNS'IIB} and \eqref{contractionRR'IIB}. This expression coincides with the one derived in \cite{Aldazabal:2010ef} on the basis of generalised geometry considerations. 
The mirror IIA superpotential is
\begin{equation}
W_{{\rm IIA/O6}}  = \int e^{J_{\rm c}} \wedge [ \mathcal{F}_{\rm RR} - \mathcal{H}_{\rm NS} \cdot \Omega_{\rm c} 
+ \mathcal{P} \cdot \Omega_{\rm c}^2 - \mathcal{H}_{\rm NS'} \cdot \Omega_{\rm c}^3 + \mathcal{F}_{\rm RR'} \cdot \Omega_{\rm c}^4 
\, ] \quad , \label{superpotIIAfull}
\end{equation}
where the contractions are defined in eqs. \eqref{contractionsNSIIA}, \eqref{contractionsPIIA}, \eqref{contractionsNS'IIA} and \eqref{contractionsRR'IIA}. 
We interpret the fact that the IIA superpotential has a unique expression when written as in 
eq. \eqref{superpotIIAfull} as a manifestation of mirror symmetry. For  geometries with $SU(3)$ structure,  $\Omega$ and $e^{J_{\rm c}}$ are two  Clifford(6,6) spinors which are both {\it pure} ({\it i.e.} they are annihilated by half of the gamma matrices) and mirror symmetry corresponds to the exchange of these two spinors \cite{Fidanza:2003zi,Grana:2005jc,Grana:2005tf,Grimm:2005fa,Grana:2006hr}. Generalising this to compactifications with non-geometric fluxes, one thus expects that the IIB superpotential $\int ({\rm flux} \cdot \mathcal{J}_{\rm c}^n  ) \wedge \Omega$ be mapped by mirror symmetry to the IIA superpotential
$ \int e^{J_{\rm c}} \wedge ({\rm flux} \cdot \Omega_{\rm c}^n )$, precisely of the form we find. 

 
 The full  expression of the superpotential as a function of the moduli in IIB is 
 \begin{align}
W_{\rm IIB/O3} & =  - \mathcal{F}_{y^1 y^2 y^3} + i U_i \mathcal{F}_{x^i y^j y^k} + U_j U_k \mathcal{F}_{y^i x^j x^k} - i U_1 U_2 U_3 \mathcal{F}_{x^1 x^2 x^3}  \nonumber \\
& - iS ( - \mathcal{H}_{y^1 y^2 y^3} + i U_i \mathcal{H}_{x^i y^j y^k} + U_j U_k \mathcal{H}_{y^i x^j x^k} - i U_1 U_2 U_3 \mathcal{H}_{x^1 x^2 x^3} )\nonumber \\
& + i T_i ( \mathcal{Q}_{y^i}^{x^j x^k} - i U_i \mathcal{Q}_{x^i}^{x^j x^k} + i U_k \mathcal{Q}_{y^i}^{x^j y^k} + i U_j \mathcal{Q}_{y^i}^{y^j x^k} + U_i U_j \mathcal{Q}_{x^i}^{y^j x^k} \nonumber \\
& + U_i U_k \mathcal{Q}_{x^i}^{x^j y^k}  - U_j U_k \mathcal{Q}_{y^i}^{y^j y^k} + i U_1 U_2 U_3 \mathcal{Q}_{x^i}^{y^j y^k} ) \nonumber \\
& + S T_i ( \mathcal{P}_{y^i}^{x^j x^k} - i U_i \mathcal{P}_{x^i}^{x^j x^k} + i U_j \mathcal{P}_{y^i}^{y^j x^k} + i U_k \mathcal{P}_{y^i}^{x^j y^k}  + U_i U_j \mathcal{P}_{x^i}^{y^j x^k} \nonumber \\
& + U_i U_k \mathcal{P}_{x^i}^{x^j y^k}  - U_j U_k \mathcal{P}_{y^i}^{y^j y^k} + i U_1 U_2 U_3 \mathcal{P}_{x^i}^{y^j y^k} )  \\
& + T_j T_k ( - \mathcal{P}^{x^i , y^i x^i x^j x^k} + i  U_i \mathcal{P}^{ y^i , x^i y^i x^j x^k} - i U_j \mathcal{P}^{ x^i , y^i x^i y^j x^k} - i U_k \mathcal{P}^{x^i , y^i x^i x^j y^k} \nonumber \\
& - U_i U_j \mathcal{P}^{y^i , x^i y^i y^j x^k} - U_i U_k \mathcal{P}^{ y^i , x^i y^i x^j y^k} + U_j U_k \mathcal{P}^{x^i, y^i x^i y^j y^k} -i U_1 U_2 U_3 \mathcal{P}^{y^i , x^i y^i y^j y^k} ) \nonumber \\
& -i S  T_j T_k ( - \mathcal{Q}'^{x^i , y^i x^i x^j x^k} + i  U_i \mathcal{Q}'^{ y^i , x^i y^i x^j x^k} - i U_j \mathcal{Q}'^{ x^i , y^i x^i y^j x^k} - i U_k \mathcal{Q}'^{x^i , y^i x^i x^j y^k} \nonumber \\
& - U_i U_j \mathcal{Q}'^{y^i , x^i y^i y^j x^k} - U_i U_k \mathcal{Q}'^{ y^i , x^i y^i x^j y^k} + U_j U_k \mathcal{Q}'^{x^i, y^i x^i y^j y^k} -i U_1 U_2 U_3 \mathcal{Q}'^{y^i , x^i y^i y^j y^k} )\nonumber \\
& + T_1 T_2 T_3 ( -i \mathcal{H}'^{x^1 x^2 x^3} + U_i \mathcal{H}'^{y^i x^j x^k} + i U_j U_k \mathcal{H}'^{x^i y^j y^k} - U_1 U_2 U_3 \mathcal{H}'^{y^1 y^2 y^3} )\nonumber \\
& -iS T_1 T_2 T_3 ( -i \mathcal{F}'^{x^1 x^2 x^3} + U_i \mathcal{F}'^{y^i x^j x^k} + i U_j U_k \mathcal{F}'^{x^i y^j y^k} - U_1 U_2 U_3 \mathcal{F}'^{y^1 y^2 y^3} ) \quad ,\nonumber 
\end{align}
where the sum over the $i$ index is understood wherever this index occurs, while $j$ and $k$ are related to $i$ by cyclicity. The mirror IIA expression is obtained by exchanging the fluxes in this expression with the IIA ones given in Tables \ref{RRNSNSPfluxestable}, \ref{allH'fluxes} and \ref{allF'fluxes} and by swapping $U_i$ with $T_i$. In the formula for simplicity of notation we have also omitted the six antisymmetric indices of $\mathcal{H}'^{3,6}$ and $\mathcal{F}'^{3,6}$, which are understood to be $x^1 x^2 x^3 y^1 y^2 y^3$.
In the next section we will focus on the tadpole cancellation conditions generated by the NS$'$ and RR$'$ fluxes and we  will determine the set of exotic branes which have to be added in order to cancel them. 
 
\section{Exotic branes, tadpoles and Bianchi identities}

In the IIB/O3 $T^6/[\mathbb{Z}_2 \times \mathbb{Z}_2]$ orientifold, D3 and D7-branes can be included in order to cancel the  tadpoles induced by the RR flux $\mathcal{F}_3$ and the NS fluxes $\mathcal{H}_3$ and $\mathcal{Q}$ \cite{Aldazabal:2006up}. 
In particular,  only stacks of D7-branes wrapping two of the internal two-dimensional tori with coordinates $x$ and $y$ are allowed. After performing  three T-dualities along the $x$ directions, one similarly finds that in the dual IIA description with O6-planes, the only allowed RR sources are D6-branes spanning three $x$ directions or one $x$ and two $y$ directions, each on a different torus.
Besides the D-branes,  more localised sources surviving the orientifold projections and compatible with supersymmetry can be included. In particular, in IIB one can include the S-dual of the D7-brane~
\cite{Aldazabal:2006up}, which is a brane with $\alpha=-3$.
In  \cite{Lombardo:2016swq} we have recently determined all the $\alpha=-3$ branes which cancel the tadpoles induced by the $\mathcal{P}$ fluxes in both the IIB and IIA orientifolds  using the universal T-duality rules in eq. \eqref{allbranesallalphasruleintro}. Except for the S-dual of the D7-brane, all these branes turn out to be exotic as they are defined only in presence of isometries, and we associate them to specific components of the mixed-symmetry potentials $E$ as reviewed in the introduction~\cite{Bergshoeff:2011ee}. 
For the convenience of the reader, we have collected all the results on the $\alpha=-1$ branes ({\it i.e.} the D-branes) and $\alpha=-3$ branes in Table \ref{CEbranestable}. 
In this section we will first determine how the NS$'$  fluxes modify the tadpole conditions for the  $\alpha=-3$ branes listed in Table \ref{CEbranestable}, and we will then 
complete the analysis of \cite{Lombardo:2016swq} determining all the exotic branes which can be included in both the type-II orientifolds in order to cancel the tadpoles induced by the NS$'$ and/or the RR$'$ fluxes.

\begin{table}[t!]
\begin{center}
\scalebox{1}{
\begin{tabular}{|c|c||c|c||c|}
\hline \multicolumn{2}{|c||}{IIB} & \multicolumn{2}{|c||}{IIA} & \# \\
 \cline{1-4} \rule[-1mm]{0mm}{2mm}  potential &  component & component & potential &  \\
 \hline \hline \rule[-2mm]{0mm}{2mm} $C_4$ & $C_{4}$ & $C_{4\, x^1 x^2 x^3}$ & $C_{7}$ &1\\
 \hline
  \rule[-2mm]{0mm}{2mm} $C_{8}$ & $C_{4\, x^i y^i x^j y^j}$ & $C_{4\,  y^i y^j x^k }$ & $C_{7}$ & 3\\ 
 \hline \hline \rule[-2mm]{0mm}{2mm} $E_8$ & $E_{4\, x^i y^i x^j y^j}$ & $E_{4\, x^i y^i x^j y^j x^k, x^i x^j x^k, x^k }$ & $E_{9,3,1}$ & 3 \\
 \hline
  \rule[-2mm]{0mm}{2mm} $E_{8,4}$ & $E_{4\, x^i y^i x^j y^j,x^i y^i x^j y^j}$ & $E_{4\,  x^i y^i x^j y^j x^k, y^i y^j x^k, x^k }$ & $E_{9,3,1}$  & 3\\ 
 
  \rule[-2mm]{0mm}{2mm} & $E_{4\, x^i y^i x^j y^k , x^i y^i x^j y^k } $ & $E_{4\, x^i y^i x^j y^k x^k , y^i y^k x^k  , x^k }$ & & 6\\
 
  \cline{2-5} \rule[-2mm]{0mm}{2mm}   & $E_{4\, x^i y^i x^j x^k , x^i y^i x^j x^k }$ & $E_{4\, x^i y^i x^j x^k, y^i}$ & $E_{8,1}$ & 3\\
  \cline{2-5} \rule[-2mm]{0mm}{2mm} & $E_{4\,  x^i y^i y^j y^k , x^i y^i y^j y^k }$ & $E_{10, y^i x^j y^j x^k y^k, x^j x^k }$ & $E_{10,5,2}$ & 3\\
  \hline  \rule[-2mm]{0mm}{2mm} $E_{9,2,1}$ & $E_{4 \, x^i y^i x^j y^j y^k, x^i y^k, x^i}$ & $E_{4 \, y^i x^j y^j x^k y^k, x^j x^k y^k , x^k}$ & $E_{9,3,1}$ & 6\\
 \rule[-2mm]{0mm}{2mm}  & $E_{4\,  x^i y^i x^j y^j x^k, y^i x^k , y^i}$ & $E_{4\,  x^i y^i x^j y^j x^k, x^i y^i x^j, y^i }$ & & 6 \\

  \cline{2-5}  \rule[-2mm]{0mm}{2mm}  & $E_{4 \, x^i y^i x^j y^j x^k, x^i x^k , x^i}$ & $E_{4\, y^i x^j y^j x^k, x^j}$ & $E_{8,1}$ & 6\\
  \cline{2-5} \rule[-2mm]{0mm}{2mm}  & $E_{4\, x^i y^i x^j y^j y^k, y^i y^k, y^i}$ & $E_{10, x^i y^i x^j x^k y^k , y^i x^k}$ & $E_{10,5,2}$ & 6\\
  \hline   \rule[-2mm]{0mm}{2mm} $E_{10,4,2}$ & $E_{10, x^i y^i x^j y^j, x^i y^i}$ & $E_{4\, y^i x^j y^j x^k y^k, y^i y^j x^k, y^i}$ & $E_{9,3,1}$ & 6\\
   \rule[-2mm]{0mm}{2mm}  & $E_{10, x^i y^j x^k y^k, x^i y^j}$ & $E_{4\, y^i x^j y^j x^k y^k, x^j y^j y^k, y^j}$ & & 6\\
  \cline{2-5} \rule[-2mm]{0mm}{2mm}  & $E_{10, x^i y^i x^j x^k, x^j x^k}$ & $E_{4\, x^i y^i y^j y^k , y^i}$ & $E_{8,1}$ & 3\\

  \cline{2-5} \rule[-2mm]{0mm}{2mm}  & $E_{10, x^i y^i y^j y^k, y^j y^k}$ & $E_{10, y^i x^j y^j x^k y^k, y^j y^k}$ & $E_{10,5,2}$  & 3\\
\hline
 \end{tabular}
}
\caption{\footnotesize The $\alpha =-1$ and  $\alpha=-3$ branes that can be included in order to cancel the tadpoles generated by the  fluxes. To avoid cumbersome notations, in all terms we have not specified the sets of ten indices. Moreover, the indices $i,j,k$ are always meant to be all different. In the last column we give the number of independent components for each potential, which corresponds to the number of different branes. There are in total 4 $\alpha=-1$ branes and 60 $\alpha=-3$ branes.}
\label{CEbranestable}
\end{center}
\end{table}

  As Table \ref{CEbranestable} shows, in IIB one has to include together with $E_8$ also the mixed-symmetry potentials $E_{8,4}$, $E_{9,2,1}$ and $E_{10,4,2}$.  The field $E_{8,4}$  electrically couples to the $3^4_3$-brane\footnote{It is conventional to denote with $p^{m}_{n}$ a   $p$-brane with $\alpha =-n$  and $m$ orthogonal isometries. The same notation was also used in  \cite{Lombardo:2016swq}.} in four dimensions. 
Using the T-duality rules of eqs. \eqref{TdualityruleNSfluxes}, \eqref{TdualityrulePfluxes} and \eqref{allbranesallalphasruleintro}, the generalised Chern-Simons term
 $\int E_8 \wedge \mathcal{P}_1^2 \cdot \mathcal{H}_3$ responsible for the tadpole of the S-dual of the D7-brane
 is mapped to a term $\int E_{8,4} \wedge (\mathcal{P}_1^2 \cdot \mathcal{Q} + \mathcal{P}^{1,4} \cdot \mathcal{H}_3)$.
Under S-duality, the  $E_{8,4}$ potential is a singlet, while the term $\mathcal{P}^{1,4} \mathcal{H}_3$ is mapped to 
$-  \mathcal{Q}'^{1,4} \mathcal{F}_3$.
 As a consequence, we find that 
when  all the allowed fluxes are taken into account the Chern-Simons coupling is
\begin{align}\label{E_{8,4}}
\int {E}_{8,4} \wedge (\mathcal{P}_1^2 \cdot \mathcal{Q}+\mathcal{P}^{1,4}\cdot \mathcal{H}_3-\mathcal{Q}'^{1,4}\cdot \mathcal{F}_3)^4_2 \quad ,
\end{align}
  which gives a tadpole condition for the $3^4_3$-brane.
In this expression, the products are defined as 
\begin{align}
&(\mathcal{P}_1^2 \cdot \mathcal{Q})^{abcd}_{ef}=12\mathcal{P}^{[ab}_{[e}\mathcal{Q}^{cd]}_{f]}\nonumber \\
&(\mathcal{P}^{1,4}\cdot \mathcal{H}_3)^{abcd}_{ef}=\mathcal{P}^{p,abcd}\mathcal{H}_{pef} \quad ,
\end{align}
and the contraction $\mathcal{Q}'^{1,4}\cdot \mathcal{H}_3$ follows by S-duality.

Besides $E_{8,4}$, in the IIB orientifold also the field $E_{9,2,1}$ is a singlet under S-duality. In four dimensions, the electric source for $E_{9,2,1}$ is the $6^{1,1}_3$-brane,\footnote{The 1,1 denotes the fact that two isometries orthogonal to the worldvolume of the brane are different; specifically, one isometry corresponds to an index repeated twice while the other one corresponds to an index repeated three times. We use similar conventions for all the branes considered in rest of this section.}  and
the corresponding Chern-Simons term takes the form
\begin{align} \label{CSE2}
\int E_{9,2,1} \wedge (\mathcal{P}_1^2 \cdot \mathcal{Q}+\mathcal{P}^{1,4}\cdot \mathcal{H}_3-\mathcal{Q}'^{1,4}\cdot \mathcal{F}_3)^{2,1}_{1} \quad ,
\end{align}
where  the  contractions between the fluxes are  defined as
\begin{align}
&(\mathcal{P}_1^2 \cdot \mathcal{Q})^{ab,a}_c=-\mathcal{P}^{ab}_p \mathcal{Q}^{pa}_c+\mathcal{Q}^{ab}_p \mathcal{P}^{pa}_c \nonumber \\
& (\mathcal{P}^{1,4}\cdot \mathcal{H}_3)^{ab,a}_c= \frac{1}{2}\mathcal{P}^{a,abpq}\mathcal{H}_{cpq} \quad ,  \label{3.7}
\end{align}
and the contraction $\mathcal{Q}'^{1,4} \cdot \mathcal{F}_3$ follows by S-duality. In  equation (4.25) of \cite{Lombardo:2016swq} the Chern-Simons term for any components of $E_{9,2,1}$, not necessarily associated to a single brane, and with only the non-geometric fluxes $\mathcal{P}_1^2$ and $\mathcal{Q}$ turned on, has been derived. One can verify that turning off the fluxes $\mathcal{P}^{1,4}$ and $\mathcal{Q}'^{1,4}$ in eq. \eqref{CSE2} reproduces the single brane components of that equation.

The last $\alpha=-3$ brane to consider in IIB is the $5_3^{2,2}$, which is the electric source for the potential is $E_{10,4,2}$, and whose tadpole condition is generated by the Chern-Simons term
\begin{equation}
\int E_{10,4,2} \times (\mathcal{P}^{1,4} \cdot \mathcal{Q}+\mathcal{H}'^{3,6}\cdot \mathcal{F}_3)^{4,2},\label{ChernSimonsE1042}
\end{equation}
with the contractions between the fluxes specified by
\begin{align}
(\mathcal{P}^{1,4}\cdot \mathcal{Q})^{abcd,cd}&=\mathcal{P}^{p,abcd}\mathcal{Q}^{cd}_{p}+\mathcal{P}^{c,cdap}\mathcal{Q}^{bd}_{p}-\mathcal{P}^{c,cdbp}\mathcal{Q}^{ad}_{p} \nonumber \\ &+\mathcal{P}^{d,dcap}\mathcal{Q}^{bc}_{p}-\mathcal{P}^{d,dcbp}\mathcal{Q}^{ac}_{p} \nonumber\\
(\mathcal{H}'^{3,6}\cdot \mathcal{F}_3)^{abcd,cd}&=-\frac{1}{2}\mathcal{H}'^{acd,abcdpq}\mathcal{F}_{apq}-\frac{1}{2}\mathcal{H}'^{bcd,abcdpq}\mathcal{F}_{bpq} \quad .\label{contractionsE1042}
\end{align}
The  potential $E_{10,4,2}$ forms a triplet of the  $SL(2,\mathbb{R})$ symmetry of the IIB theory, and in particular under S-duality it is mapped to  $G_{10,4,2}$ which in turn sources the  $5^{2,2}_{5}$-brane with $\alpha=-5$. Transforming under S-duality eq. \eqref{ChernSimonsE1042}, one obtains  that the tadpole condition for this brane  is generated by the Chern-Simons term
\begin{equation}
\int G_{10,4,2} \times (\mathcal{Q}'^{1,4} \cdot \mathcal{P}_1^2+\mathcal{F}'^{3,6}\cdot \mathcal{H}_3)^{4,2}, \label{ChernSimonsG1042}
\end{equation}
where the contractions are defined in a way analogous to eq. \eqref{contractionsE1042}.  We now want to determine all the  $\alpha=-5$ branes that can be 
simultaneously included in the IIB theory together with the $5^{2,2}_{5}$-brane. 
 
In the orientifold model, the components of the $E_{10,4,2}$ potential that give rise to branes in IIB are given on the left-hand side of the last four rows in Table \ref{CEbranestable}. By S-duality, these are mapped to the same components of the $G_{10,4,2}$ potential. Using eq. \eqref{allbranesallalphasruleintro} with $\alpha=-5$, one discovers that by performing three T-dualities along the $x$ directions these components are mapped to components of $G_{10,5,3,1}$, $G_{10,4,1}$ and $G_{10,6,5,2}$ in IIA. In particular,  the IIA orientifold condition implies that  the number of $y$ indices must be even. The additional components of these potentials that satisfy these conditions are mapped by mirror symmetry to the components of the IIB potentials $G_{10,5,4,1}$, $G_{10,6,2,2}$ and $G_{10,6,6,2}$, corresponding to  $4_{5}^{1,3,1}$, $3_{5}^{4,0,2}$ and $3_{5}^{0,4,2}$ branes. The full list of $\alpha=-5$ branes allowed in the orientifold, for both the IIB and IIA case, is given in Table \ref{Gbranestable}.

\begin{table}[t!]
\begin{center}
\scalebox{1.05}{
\begin{tabular}{|c|c||c|c||c|}
\hline \multicolumn{2}{|c||}{IIB} & \multicolumn{2}{|c||}{IIA}  & \#\\
 \cline{1-4} \rule[-1mm]{0mm}{2mm}  potential &  component & component & potential & \\
\hline
  \hline   \rule[-2mm]{0mm}{2mm} $G_{10,4,2}$ & $G_{10, x^i y^i x^j y^j, x^i y^i}$ & $G_{10 , x^iy^ix^jy^jx^k,x^jy^ix^k, x^k}$ & $G_{10,5,3,1}$ & 6\\
   \rule[-2mm]{0mm}{2mm}  & $G_{10, x^i y^j x^k y^k, x^i y^j}$ & $G_{10, x^i y^j x^k y^k x^j, x^j y^j x^k, x^j}$ &  & 6\\
  \cline{2-5} \rule[-2mm]{0mm}{2mm}  & $G_{10, x^i y^i x^j x^k, x^j x^k}$ & $G_{10 , x^i y^i x^k x^j, x^i}$ & $G_{10,4,1}$ & 3\\

  \cline{2-5} \rule[-2mm]{0mm}{2mm}  & $G_{10, x^i y^i y^j y^k, y^j y^k}$ & $G_{10, 6, x^i x^j x^k y^j y^k, x^j x^k}$ & $G_{10,6,5,2}$ & 3\\

 \hline   \rule[-2mm]{0mm}{2mm} $G_{10,5,4,1}$ & $G_{10, x^i y^i x^j y^j x^k, x^i y^i x^j x^k, x^j}$ & $G_{10 , x^i y^i x^k y^j, y^i}$ & $G_{10,4,1}$ & 6\\
  \cline{2-5} \rule[-2mm]{0mm}{2mm}  & $G_{10, x^i y^i x^j y^j x^k, x^j y^j y^i x^k, y^i}$ & $G_{10, x^i y^i x^j y^j x^k, x^i y^i y^j, y^i}$ & $G_{10,5,3,1}$ & 6 \\
 \rule[-2mm]{0mm}{2mm}  & $G_{10, x^k y^k x^j y^j y^i ,x^j y^j x^k y^i, x^k}$ & $G_{10 , x^i y^i x^j y^j y^k, x^i y^i y^j, x^i}$ &  & 6\\

  \cline{2-5} \rule[-2mm]{0mm}{2mm}  & $G_{10, x^j y^j x^k y^k y^i, x^k y^k y^i y^j, y^j}$ & $G_{10, 6 , x^i y^i x^j y^j y^k, y^j x^i}$ & $G_{10,6,5,2}$ & 6\\

 \hline   \rule[-2mm]{0mm}{2mm} $G_{10,6,2,2}$ & $G_{10, 6, x^i y^i, x^i y^i}$ & $G_{10 , x^j y^j x^k y^k y^i , y^i x^j x^k, y^i}$ & $G_{10,5,3,1}$ & 3 \\
   \rule[-2mm]{0mm}{2mm}  &  $G_{10, 6, x^i y^j, x^i y^j}$ & $G_{10 , x^j y^j x^k y^k y^i, y^j x^j x^k, y^j}$ & & 6\\
\cline{2-5}  \rule[-2mm]{0mm}{2mm}  &  $G_{10, 6, y^i y^j, y^i y^j}$ & $G_{10, 6, x^i y^i x^j y^j x^k, y^j y^i}$ & $G_{10,6,5,2}$ & 3 \\

  \cline{2-5} \rule[-2mm]{0mm}{2mm}  & $G_{10, 6, x^i x^j, x^i x^j}$ & $G_{10, y^j y^i x^k y^k, x^k}$ & $G_{10,4,1}$ & 3 \\

\hline   \rule[-2mm]{0mm}{2mm} $G_{10,6,6,2}$ & $G_{10, 6, 6, x^i y^i}$ & $G_{10 , x^j y^j x^k y^k y^i , y^j y^i y^k, y^i}$ & $G_{10,5,3,1}$ & 3 \\
 
\hline

 \end{tabular}
}

\caption{\footnotesize The $\alpha=-5$ branes that can be included in order to cancel the tadpoles generated always by the NS$'$ and RR$'$ fluxes. To avoid cumbersome notations, in all terms we have not specified the sets of ten and six internal indices. Moreover, the indices $i$, $j$, $k$ are always understood to be different. In the last column we give the number of independent components for each potential, which corresponds to the number of different branes, which are 60 in total.}
\label{Gbranestable}
\end{center}
\end{table}

We now proceed with discussing the tadpole conditions for the remaining $\alpha=-5$ branes in the IIB theory, which we obtain precisely with the method discussed above. We first map the relevant brane components in eq. \eqref{ChernSimonsG1042} to the mirror IIA theory, we then extend this result to the remaining branes in the IIA theory that are allowed and we finally map this back to IIB. 
In particular, for $G_{10,5,4,1}$ we find the term
\begin{equation} \label{G1054}
 \int G_{10,5,4,1} \times (\mathcal{P}_1^2 \cdot \mathcal{H}'^{3,6}-\mathcal{Q} \cdot \mathcal{F}'^{3,6}+\mathcal{P}^{1,4}\cdot \mathcal{Q}'^{1,4})^{5,4,1} \quad ,
\end{equation}
where we have defined
\begin{align}
(\mathcal{P}_1^2 \cdot \mathcal{H}'^{3,6})^{abcde,abcd,a}=&-\mathcal{P}_p^{ad}\mathcal{H}'^{abc,abcdep}+\mathcal{P}^{ac}_p\mathcal{H}'^{abd,abcdep}-\mathcal{P}^{ab}_p \mathcal{H}'^{acd,abcdep} \nonumber \\
(\mathcal{P}^{1,4}\cdot \mathcal{Q}'^{1,4})^{abcde,abcd,a}= &\mathcal{P}^{a,acde}\mathcal{Q}'^{b,bcda}-\mathcal{P}^{b,bcda}\mathcal{Q}'^{a,acde}+\mathcal{P}^{a,abde}\mathcal{Q}'^{c,cdab} \\  \nonumber
&-\mathcal{P}^{c,cdab}\mathcal{Q}'^{a,abde}+\mathcal{P}^{a,abce}\mathcal{Q}'^{d,dabc}-\mathcal{P}^{d,dabc}\mathcal{Q}'^{a,abce} 
\end{align}
and $(\mathcal{Q} \cdot \mathcal{F}'^{3,6})^{5,4,1}$ is found by S-duality. Moreover, from the form of the Chern-Simons coupling in eq. \eqref{G1054}, one can see that $G_{10,5,4,1}$ is a singlet under S-duality. The field $G_{10,6,2,2}$ is also a singlet and, by the same analysis, its generalised Chern-Simons term is found to be
\begin{equation} 
\int G_{10,6,2,2} \times (\mathcal{P}^{1,4} \cdot \mathcal{Q}'^{1,4}+\mathcal{P}_1^2 \cdot \mathcal{H}'^{3,6}-\mathcal{Q} \cdot \mathcal{F}'^{3,6})^{6,2,2} \quad ,\label{ChernSimonsG10622}
\end{equation}
with 
\begin{align}
(\mathcal{P}^{1,4}\cdot \mathcal{Q}'^{1,4})^{abcdef,ab,ab}&=-\mathcal{P}^{a,abcd}\mathcal{Q}'^{b,befa}-\mathcal{P}^{a,abef}\mathcal{Q}'^{b,bcda}+\mathcal{P}^{a,abce}\mathcal{Q}'^{b,bdfa}-\mathcal{P}^{a,abcf} \nonumber \mathcal{Q}'^{b,bdea}\\ &-\mathcal{P}^{a,abde}\mathcal{Q}'^{b,bcfa}+\mathcal{P}^{a,abdf}\mathcal{Q}'^{b,bcea}+\mathcal{P}^{b,bcda}\mathcal{Q}'^{a,abef}+\mathcal{P}^{b,befa}\mathcal{Q}'^{a,abcd}\nonumber \\ 
&-\mathcal{P}^{b,bcea}\mathcal{Q}'^{a,abdf}+\mathcal{P}^{b,bcfa}\mathcal{Q}'^{a,abde}+\mathcal{P}^{b,bdea}\mathcal{Q}'^{a,abcf}-\mathcal{P}^{b,bdfa}\mathcal{Q}'^{a,abce}\nonumber \\
(\mathcal{P}_1^2 \cdot \mathcal{H}'^{3,6})^{abcdef,ab,ab}=&-\mathcal{P}^{ab}_p \mathcal{H}'^{abp,abcdef} \quad .
\end{align}
The remaining IIB $\alpha=-5$ field in Table \ref{Gbranestable} is $G_{10,6,6,2}$ and the corresponding Chern-Simons coupling is
\begin{equation} \label{ChernSimonsG10662}
 \int G_{10,6,6,2} \times (\mathcal{P}^{1,4} \cdot \mathcal{H}'^{3,6})^{6,6,2} \quad ,
\end{equation}
where we have defined
\begin{align}
&(\mathcal{P}^{1,4}\cdot \mathcal{H}'^{3,6})^{abcdef,abcdef,ab}= \nonumber \\ 
&\mathcal{P}^{a,abef}\mathcal{H}'^{bcd,abcdef}-\mathcal{P}^{a,abdf}\mathcal{H}'^{bce,abcdef}+\mathcal{P}^{a,abde}\mathcal{H}'^{bcf,abcdef}+\mathcal{P}^{a,abcf}\mathcal{H}'^{bde,abcdef} \nonumber \\ 
&-\mathcal{P}^{a,abce}\mathcal{H}'^{bdf,abcdef}+\mathcal{P}^{a,abcd}\mathcal{H}'^{bef,abcdef}+\mathcal{P}^{b,befa}\mathcal{H}'^{acd,abcdef}-\mathcal{P}^{b,bdfa}\mathcal{H}'^{ace,abcdef}  \nonumber \\
&+\mathcal{P}^{b,bdea}\mathcal{H}'^{acf,abcdef}+\mathcal{P}^{b,bcfa}\mathcal{H}'^{ade,abcdef}-\mathcal{P}^{b,bcea}\mathcal{H}'^{adf,abcdef} + \mathcal{P}^{b,bcda}\mathcal{H}'^{aef,abcdef} \! .\label{contractionsG10662}
\end{align}

\begin{table}[t!]
\begin{center}
\scalebox{1.05}{
\begin{tabular}{|c|c||c|c||c|}
\hline \multicolumn{2}{|c||}{IIB} & \multicolumn{2}{|c||}{IIA} &  \#\\
 \cline{1-4} \rule[-1mm]{0mm}{2mm}  potential &  component & component & potential & \\

 \hline \hline   \rule[-2mm]{0mm}{2mm} $I_{10,6,6,2}$ & $I_{10, 6, 6, x^i y^i}$ & $I_{10, 6 , 6, x^j y^i x^k}$ & $I_{10,6,6,3}$ & 3\\
  \hline   \rule[-2mm]{0mm}{2mm} $I_{10,6,6,6}$  & $I_{10, 6, 6, 6}$ & $I_{10,6,6, y^1 y^2 y^3}$ & $I_{10,6,6,3}$ & 1\\

\hline

 \end{tabular}
}

\caption{\footnotesize The $\alpha=-7$ branes that can be included in order to cancel the tadpoles generated by the NS$'$ and RR$'$ fluxes. In order to avoid cumbersome notations, in all terms we have not specified the sets of ten and six internal indices.  In the last column we give the number of independent components for each potential, which corresponds to the number of different branes, which are 4 in total.}
\label{Ibranestable}
\end{center}
\end{table}

The field $G_{10,6,6,2}$ transforms in a triplet of the IIB $SL(2,\mathbb{R})$ symmetry, and in particular by S-duality it is mapped to $I_{10,6,6,2}$, which is associated to the $3_{7}^{0,4,2}$-brane, which has $\alpha=-7$. Thus, one can conjecture that also the $\alpha=-7$ branes electrically coupled to $I_{10,6,6,2}$ can be included in IIB in order to cancel other tadpoles induced by the RR$'$ and NS$'$ fluxes. The exotic branes which couple electrically to $I_{10,6,6,2}$ can wrap only one of the two-dimensional tori of the compactification and under mirror-symmetry, using the rules \eqref{allbranesallalphasruleintro}, they are found to be in correspondence with three of the four $3_{7}^{0,3,3}$-branes associated to the field $I_{10,6,6,3}$ in IIA, in particular those components with two $x$ and one $y$ as the extra indices. Once again, one can argue that also the remaining brane associated with the component of $I_{10,6,6,3}$ with three extra $y$ can be included in IIA and this in turn corresponds to the $3_{7}^{0,0,6}$-brane associated to $I_{10,6,6,6}$ in IIB. We have collected all the results for the $\alpha=-7$ branes in Table \ref{Ibranestable}.

The Chern-Simons of the $I_{10,6,6,2}$  potential in IIB is obtained from S-duality on the one of $G_{10,6,6,2}$ in eq. \eqref{ChernSimonsG10662}, resulting in
 \begin{equation}
\int I_{10,6,6,2} \times (\mathcal{Q}'^{1,4} \cdot \mathcal{F}'^{3,6})^{6,6,2} \quad ,
\end{equation}
where the contractions are analogous to eq. \eqref{contractionsG10662}.
The Chern-Simons term associated to the field $I_{10,6,6,6}$ is reminiscent of that induced by the D3-brane charge and proportional to $\mathcal{H}_3 \wedge \mathcal{F}_3$. Namely, we find
\begin{equation} 
 \int I_{10,6,6,6} \times (\mathcal{H}'^{3,6} \cdot \mathcal{F}'^{3,6})^{6,6,6} \quad ,\label{ChernSimonsI10666}
\end{equation}
and the contraction between the fluxes is specified by
\begin{align}
(\mathcal{H}'^{3,6} \cdot \mathcal{F}'^{3,6})^{abcdef,abcdef,abcdef}=-20\mathcal{H}'^{[abc|,abcdef}\mathcal{F}'^{|def],abcdef} \quad .
\end{align}
Given that the potential $I_{10,6,6,6}$ is invariant under S-duality, this concludes the analysis of tadpole conditions.

In \cite{Lombardo:2016swq} we have also determined the most general Bianchi identities corresponding to the absence of $\alpha=-2$ solitonic branes (see eq. (4.2) of that paper). In the IIB orientifold only two classes of those constraints survive the orientifold projections meaning that, for the components corresponding to single branes, neither KK-monopoles (which are $5_2^1$-branes) nor  non-geometric $5_2^3$-branes can be included in the four-dimensional theory because they would break the $\mathcal{N}=1$ supersymmetry. 
Moreover, in IIA one finds that the mirror of these constraints are sourced again by KK-monopoles and $5_2^3$-branes. 
By T-duality, we are now able to determine similar constraints related to the absence of $\alpha=-4$ and $\alpha=-6$ branes in both the orientifolds. For the convenience of the reader, we have schematically anticipated the final results in eq. \eqref{Bianchibranesintro}. First, let us point out that the $5_2^3$-brane sources the potential $D_{9,3}$ in the four-dimensional theory, which actually belongs to a triplet of the IIB $SL(2,\mathbb{R})$ symmetry and in particular it is mapped to  $F_{9,3}$ by S-duality. 
Actually, for the field $F_{9,3}$ only two kinds of constraints are allowed. These are conditions on the absence of branes associated to the components with the three isometric indices containing both the coordinates $x$ and $y$ of one of the two-dimensional sub-tori.   
 By using the T-duality rules in eq. \eqref{allbranesallalphasruleintro} for the $\alpha=-4$ fields, we find that the conditions given by the field $F_{9,3}$ in IIB, corresponds to requiring the absence of specific components of the fields $F_{9,4,1}$ and $F_{10,5,2,1}$ in IIA. The requirement that in IIA  all the components of these fields with an odd number of $y$ indices have to be included implies that other six conditions on the fluxes must be imposed. In IIB, the mirror of these constraints corresponds to forbidding the exotic branes $3_{4}^{3,2}$, $5_{4}^{3,0,1}$, $3_{4}^{3,2,1}$,  sourcing respectively the fields $F_{9,5,2}$, $F_{10,4,1,1}$ and $F_{10,6,3,1}$. In all, the Bianchi identities for the IIB $\alpha=-4$ branes are found to be
  \begin{align} \label{BI}
 &3\mathcal{P}^{[ab}_p \mathcal{P}^{c]p}_f+\mathcal{Q}'^{a,abcp}\mathcal{H}_{fap}+\mathcal{Q}'^{b,abcp}\mathcal{H}_{fbp}+\mathcal{Q}'^{c,abcp}\mathcal{H}_{fcp}=0 \nonumber \\ \vspace{0.8 cm}
&-\mathcal{P}^{d,dbcp}\mathcal{P}^{ad}_p+\mathcal{P}^{d,dacp}\mathcal{P}^{bd}_p-\mathcal{P}^{d,dabp}\mathcal{P}^{cd}_p+\mathcal{Q}'^{d,dbcp}\mathcal{Q}^{ad}_p-\mathcal{Q}'^{d,dacp}\mathcal{Q}^{bd}_p+\mathcal{Q}'^{d,dabp}\mathcal{Q}^{cd}_p=0 \nonumber \\
&-\mathcal{P}^{a,abde}\mathcal{P}^{ce}_f+\mathcal{P}^{e,eabd}\mathcal{P}^{ac}_f+\mathcal{P}^{a,acde}\mathcal{P}^{be}_f-\mathcal{P}^{e,eacd}\mathcal{P}^{ab}_f-\mathcal{P}^{a,abce}\mathcal{P}^{ed}_f-\mathcal{P}^{e,eabc}\mathcal{P}^{ad}_f+\mathcal{Q}'^{a,abde}\mathcal{Q}^{ce}_f \nonumber \\
&-\mathcal{Q}'^{e,eabd}\mathcal{Q}^{ac}_f-\mathcal{Q}'^{a,acde}\mathcal{Q}^{be}_f+\mathcal{Q}'^{e,eacd}\mathcal{Q}^{ab}_f-\mathcal{Q}'^{a,abce}\mathcal{Q}^{de}_f=0\nonumber \\
&\mathcal{P}^{f,fcde}\mathcal{P}^{a,abef}-\mathcal{P}^{f,facd}\mathcal{P}^{e,efab}+\mathcal{P}^{f,fabd}\mathcal{P}^{e,efac}-\mathcal{P}^{f,fbde}\mathcal{P}^{a,acef}+\mathcal{P}^{f,fbce}\mathcal{P}^{a,adef}\nonumber \\&-\mathcal{P}^{f,fabc}\mathcal{P}^{e,efad}-\mathcal{H}'^{acf,abcdef}\mathcal{Q}^{ef}_c-\mathcal{H}'^{adf,abcdef}\mathcal{Q}^{ef}_d-\mathcal{H}'^{abf,abcdef}\mathcal{Q}^{ef}_b=0 \quad .
\end{align}

 \begin{table}[h]
\begin{center}
\scalebox{1.05}{
\begin{tabular}{|c|c||c|c||c|}
\hline \multicolumn{2}{|c||}{IIB} & \multicolumn{2}{|c|}{IIA} &  \#\\
 \cline{1-4} \rule[-1mm]{0mm}{2mm}  potential &  component & component & potential & \\
 \hline \hline \rule[-2mm]{0mm}{2mm} $D_{7,1}$ & $D_{4\, x^j y^j x^k, x^k}$ & $D_{4\, x^j y^j x^i, x^i}$ & $D_{7,1}$ & 6\\
  \rule[-2mm]{0mm}{2mm}   & $D_{4\, x^j y^j y^k, y^k }$ & $D_{4\, x^j y^j y^k x^i x^k, y^k x^i x^k }$ & $D_{9,3}$ & 6\\ 
 \hline
  \rule[-2mm]{0mm}{2mm} $D_{9,3}$ & $D_{4\, x^iy^i x^k y^k x^j, x^i y^i x^j}$ & $D_{4\, y^i x^k y^k, y^i}$ & $D_{7,1}$ & 6 \\
 
 \rule[-2mm]{0mm}{2mm}   & $D_{4\, x^iy^i x^k y^k y^j, x^i y^i y^j}$ & $D_{4\, y^i x^k y^k y^j x^j, y^i y^j x^j}$ & $D_{9,3}$& 6 \\
 
  \hline \hline \rule[-2mm]{0mm}{2mm} $F_{9,3}$ & $F_{4\, x^iy^i x^k y^k x^j, x^i y^i x^j}$ & $F_{4\, x^iy^i x^k y^k x^j, x^i y^i x^jx^k, x^k}$ & $F_{9,4,1}$ & 6\\
  
  \rule[-2mm]{0mm}{2mm}  & $F_{4\, x^iy^i x^k y^k y^j, x^i y^i y^j}$ & $F_{10, x^i y^i y^jx^jx^k, x^jx^k, x^j}$ & $F_{10,5,2,1}$ & 6\\
    \hline
 
 \rule[-2mm]{0mm}{2mm} $F_{9,5,2}$  & $F_{4\, x^iy^ix^jy^jx^k, x^iy^ix^jy^jx^k,x^jy^j}$ & $F_{4\, x^iy^ix^jy^jx^k, x^iy^iy^jx^k,y^j}$ & $F_{9,4,1}$ & 6\\
  \rule[-2mm]{0mm}{2mm} & $F_{4\, x^iy^ix^jy^jy^k, x^iy^ix^jy^jy^k, x^jy^j}$ & $F_{10, x^iy^iy^jy^kx^k, y^jx^k,x^k}$ & $F_{10,5,2,1}$ & 6\\
    \hline
    
    \rule[-2mm]{0mm}{2mm} $F_{10,4,1,1}$  & $F_{10,x^ky^kx^iy^i,x^k, x^k}$ & $F_{4\, x^iy^i x^j y^j y^k, y^kx^iy^ix^j, x^j}$ & $F_{9,4,1}$ & 6\\
  \rule[-2mm]{0mm}{2mm} &  $F_{10,x^ky^kx^iy^i,y^k, y^k}$ & $F_{10,x^ky^kx^iy^ix^j,y^kx^j, y^k}$ & $F_{10,5,2,1}$ & 6\\
    \hline
    
    \rule[-2mm]{0mm}{2mm} $F_{10,6,3,1}$  & $F_{10,6,x^ky^kx^j, x^j}$ &  $F_{4\,x^iy^iy^jx^ky^k,x^iy^iy^jy^k,y^k}$ & $F_{9,4,1}$ & 6\\
  \rule[-2mm]{0mm}{2mm} &  $F_{10,6,x^ky^ky^j, y^j}$ & $F_{10, x^iy^ix^jy^jy^k, y^ky^j, y^j}$    & $F_{10,5,2,1}$ & 6\\
    \hline \hline
    
     \rule[-2mm]{0mm}{2mm} $H_{10,6,3,1}$ & $H_{10,6,x^ky^kx^j, x^j}$ & $H_{10, 6, x^ky^kx^i, x^i}$  & $H_{10,6,3,1}$ & 6\\
  
  \rule[-2mm]{0mm}{2mm}  & $H_{10,6,x^ky^ky^j, y^j}$ & $H_{10,6,x^ky^ky^jx^ix^j, y^jx^ix^j}$ & $H_{10,6,5,3}$ & 6\\
    \hline
 
 \rule[-2mm]{0mm}{2mm} $H_{10,6,5,3}$  & $H_{10,6,x^iy^ix^jy^jx^k,x^iy^ix^k}$ & $H_{10,6,y^ix^jy^j,y^i}$ & $H_{10,6,3,1}$ & 6\\
  \rule[-2mm]{0mm}{2mm} & $H_{10,6,x^iy^ix^jy^jy^k,x^iy^iy^k}$ & $H_{10,6,y^ix^jy^jy^kx^k,y^iy^kx^k}$ & $H_{10,6,5,3}$ & 6 \\
    \hline

 \end{tabular}
}
\caption{\footnotesize The branes with $\alpha$ even that are forbidden by the $\mathcal{N}=1$ supersymmetry and give rise to non-trivial Bianchi identities. The total number of such branes is 96.}
\label{alphaevenbranestable}
\end{center}
\end{table}

 As far as S-duality in concerned, the fields $F_{9,5,2}$ and $F_{10,4,1,1}$ turn out to be singlets, whereas $F_{10,6,3,1}$ transforms in a triplet and  its S-dual is the potential $H_{10,6,3,1}$, which has $\alpha=-6$. Exactly as in the case of the $\alpha=-2$ branes, one finds that the corresponding IIB constraint is mapped by mirror symmetry to constraints associated to specific components of the potentials $H_{10,6,3,1}$ and $H_{10,6,5,3}$. The additional IIA components of these potentials giving non-trivial constraints are then mapped in IIB to constraints which correspond to forbidding the $3_{6}^{1,2,3}$-branes which source $H_{10,6,5,3}$. 
In all, for the IIB $\alpha=-6$ branes, we find the Bianchi identities 
\begin{align} \label{BI2}
& \mathcal{Q}'^{f,fcde}\mathcal{Q}'^{a,abef}-\mathcal{Q}'^{f,facd}\mathcal{Q}'^{e,efab}+\mathcal{Q}'^{f,fabd}\mathcal{Q}'^{e,efac}-\mathcal{Q}'^{f,fbde}\mathcal{Q}'^{a,acef}+\mathcal{Q}'^{f,fbce}\mathcal{Q}'^{a,adef}\nonumber \\&-\mathcal{Q}'^{f,fabc}\mathcal{Q}'^{e,efad}-\mathcal{F}'^{acf,abcdef}\mathcal{P}^{ef}_c-\mathcal{F}'^{adf,abcdef}\mathcal{P}^{ef}_d-\mathcal{F}'^{abf,abcdef}\mathcal{P}^{ef}_b=0
\end{align}
and 
\begin{align}
&-\mathcal{F}'^{abc,abcdef}\mathcal{P}^{f,fbce}+\mathcal{F}'^{bce,abcdef}\mathcal{P}^{f,fabc} +\mathcal{F}'^{acf,abcdef}\mathcal{P}^{b,bcef}-\mathcal{F}'^{cef,abcdef}\mathcal{P}^{b,bcfa}\nonumber \\
&+\mathcal{F}'^{abf,abcdef}\mathcal{P}^{c,cefb}-\mathcal{F}'^{efb,abcdef}\mathcal{P}^{c,cfab} +\mathcal{H}'^{abc,abcdef} \mathcal{Q}'^{f,fbce} \nonumber \\ 
&-\mathcal{H}'^{bce,abcdef}\mathcal{Q}'^{f,fabc}+\mathcal{H}'^{cef,abcdef}\mathcal{Q}'^{b,bcfa}-\mathcal{H}'^{acf,abcdef}\mathcal{Q}'^{b,bcef}\nonumber \\
&-\mathcal{H}'^{abf,abcdef}\mathcal{Q}'^{c,cefb}+\mathcal{H}'^{efb,abcdef}\mathcal{Q}'^{c,cfab}=0 \quad .\label{BI22}
\end{align}

In Table \ref{alphaevenbranestable} we have collected all the branes with $\alpha$ even, forbidden by the $\mathcal{N}=1$ supersymmetry and giving rise to non-trivial Bianchi identities.
As far as the Bianchi identities are concerned, what actually happens in IIA can be easily found from \eqref{BI}, \eqref{BI2} and \eqref{BI22} by mirror symmetry. The same is true for all the IIB tadpole conditions derived in this section. In both the type II orientifolds, $2^{7}$ branes can be included in order to cancel the tadpoles induced by the fluxes. It's interesting to note that the total number of branes  compatible with the minimal amount of supersymmetry turns out to be equal to that of the fluxes. In  \cite{Aldazabal:2006up} all the fluxes have been embedded  in the representation \textbf{(2,2,2,2,2,2,2)} of $SL(2,\mathbb{R})^{7}$. In the next section we will show how also the $2^{7}$ branes of this section organise in irreducible representations of $SL(2,\mathbb{R})^{7}$ which are singlets and triplets of each $SL(2,\mathbb{R})$. An analogous result will be found for the forbidden  branes with $\alpha$ even giving rise to the Bianchi identities \eqref{BI}, \eqref{BI2} and \eqref{BI22}.

\section{Exotic branes and $SL(2,\mathbb{Z})^7$}

The orientifold model considered in this paper has a conjectured  $SL(2,\mathbb{Z})^7$ non-perturbative duality symmetry \cite{Aldazabal:2006up,Aldazabal:2008zza}, where each $SL(2,\mathbb{Z})$ has a modular action 
\begin{equation}
M \rightarrow \frac{aM-ib}{icM+d} \label{modulartransf}
\end{equation}
on each modulus $M$. We denote with $SL(2,\mathbb{R})|_{S}$, $SL(2,\mathbb{R})|_{T_i}$ and $SL(2,\mathbb{R})|_{U_i}$ the group that acts on the modulus $S$, $T_i$ and $U_i$ respectively.
The 128 fluxes that can be turned on in the model belong to the $({\bf 2,2,2,2,2,2,2})$ representation of $SL(2,\mathbb{R})^7$ \cite{Aldazabal:2006up}, where the first doublet is with respect to  $SL(2,\mathbb{R})|_{S}$, the second to fourth with respect to $SL(2,\mathbb{R})|_{T_i}$ and the fifth to seventh with respect to $SL(2,\mathbb{R})|_{U_i}$.
The aim of this section is to determine the representations of all the space-filling branes that can be included to cancel the tadpoles. 
We  will find that the branes belong to 16 irreducible representations, each made of three triplets and four singlets. More precisely, these 16 representations are all the possible representations made of three triplets and four singlets with the condition that in the IIB picture there is an even number of triplets with respect to the groups $SL(2,\mathbb{R})|_{U_i}$. Additionally, we will find that the branes with even $\alpha$, that by being projected out give rise to  Bianchi identities, collect in 12 irreducible representations of $SL(2,\mathbb{R})^7$, which again are made of three triplets and four singlets.

The analysis of \cite{Aldazabal:2006up} consisted in identifying the $SL(2,\mathbb{R})^7$ weights of each flux. The outcome can be found for instance in table 6 of that paper, where $+$ and $-$ denote the two weights of the ${\bf 2}$ of each $SL(2,\mathbb{R})$. We are interested in particular in the transformation with $b=-1$ and $c=1$ in eq. \eqref{modulartransf}, whose action on the modulus $M$ is
\begin{equation}
M \rightarrow 1/M \quad .\label{modulusinversion}
\end{equation}
This transforms a particular component of the $({\bf 2,2,2,2,2,2,2})$ representation according to  \cite{Aldazabal:2006up}
\begin{equation} \label{fluxesweightstransf}
(n_1,..., n_{M},..., n_7) \rightarrow \text{sign}(n_M) (n_1,...,-n_{M},...,n_{7}) \quad , 
\end{equation}
where   any label $n$ is either $+$ or $-$.  
We denote with {S}, $\mathcal{T}_i$ and $\mathcal{U}_i$ the generators that invert respectively the moduli $S$, $T_i$ and $U_i$ as in eq. \eqref{modulusinversion}.\footnote{In IIB, the {S} transformation is precisely S-duality, acting as $S \rightarrow 1/S$ on the axion-dilaton.}   
In the IIB/O3 orientifold, the flux $\mathcal{F}_{y^1 y^2 y^3}$ has weight $(+,+,+,+,+,+,+)$. This is mapped under {S} to the flux $\mathcal{H}_{y^1 y^2 y^3}$ of weight $(-,+,+,+,+,+,+)$. The action of the generator $\mathcal{U}_i$ on the flux components corresponds to the exchange of the coordinates $x^i$ and  $y^i$. Therefore, the fluxes  $\mathcal{F}_{x^1 x^2 x^3}$ and $\mathcal{H}_{x^1 x^2 x^3}$ have weight $(+,+,+,+,-,-,-)$ and $(-,+,+,+,-,-,-)$ respectively. The generator $\mathcal{T}_i$ maps for instance $\mathcal{F}_{y^1 y^2 y^3}$ to $\mathcal{Q}_{y^i}^{x^j x^k}$ and $\mathcal{F}_{x^1 x^2 x^3}$ to $\mathcal{Q}_{x^i}^{y^j y^k}$. In particular, the flux $\mathcal{Q}_{y^1}^{x^2 x^3}$ has weight 
$(+,-,+,+,+,+,+)$. Similarly, one can determine the weights of all the other fluxes \cite{Aldazabal:2006up}.

We now want to determine the representations of the 128 branes that can be introduced in the model to cancel the tadpoles induced by the fluxes, as discussed in the previous section. Given that the charges of the branes  have to cancel quadratic terms in the fluxes, the branes must organise themselves into representations which are either singlets or triplets of each of the seven $SL(2,\mathbb{R})$'s. 
The key idea to determine these representations is to trace the transformations of the potentials which couple to the branes from those of the fluxes. Again, we focus on the IIB/O3 orientifold, and by transforming the fluxes in  the generalised Chern-Simons terms derived in the previous section under {S}, $\mathcal{T}_i$ and $\mathcal{U}_i$ we manage to determine all the representations of the branes.

We first consider the D3-branes. 
From the fact that the RR field $C_4$ couples to the background fluxes as $C_4 \wedge \mathcal{H}_3 \wedge \mathcal{F}_3$ in the corresponding Chern-Simons term, one finds that $C_4$ does not transform under {S} and $\mathcal{U}_i$, while under the action of $\mathcal{T}_i$ it is mapped to the component $E_{{4}\, x^jy^jx^ky^k,x^jy^jx^ky^k}$\,\footnote{Using the notation of the previous section, we take the indices $i$, $j$ and $k$ to be all different.}
 of the field $E_{8,4}$. To complete the representation, one has to determine how the generators $\mathcal{T}_j$ and $\mathcal{T}_k$ act on $E_{{4}\, x^jy^jx^ky^k,x^jy^jx^ky^k}$. From \eqref{E_{8,4}}, one can check that they generate the component $G_{10,6,x^ky^k,x^ky^k}$ of $G_{10,6,2,2}$ and the field $I_{10,6,6,6}$ whose Chern-Simons terms have been given in \eqref{ChernSimonsG10622} and \eqref{ChernSimonsI10666}. Thus, we find that, starting from the RR potential $C_4$, there is a total of eight branes which can be reached by acting with the generators of $SL(2,\mathbb{R})^{7}$. These eight branes correspond to the products of the two roots of each $SL(2,\mathbb{R})|_{T_i}$, and therefore the resulting representation is ${\bf (1,3,3,3,1,1,1)}$, {\it i.e.} a singlet under $SL(2,\mathbb{R})|_{S}$ and $SL(2,\mathbb{R})|_{U_i}$ and a triplet under each  $SL(2,\mathbb{R})|_{T_i}$. 
We denote the resulting representation as $({\bf 3}_{T_1} , {\bf 3}_{T_2}, {\bf 3}_{T_3}  )$. The result is summarised in the first row of Table \ref{oddbranesB}.

 \begin{table}[h!]
\begin{center}
\scalebox{1}{
\begin{tabular}{|c|p{1.5cm} p{4.3cm}|c|c|}
\hline
\rule[3 mm]{0 mm}{2 mm} IIB/O3 $SL(2,\mathbb{R})^7$ rep & potential & component & \# & \# reps \\
\hline
\hline
\rule[3 mm]{0 mm}{1 mm}\multirow{4}{*}{$({\bf 3}_{T_1} , {\bf 3}_{T_2}, {\bf 3}_{T_3}  )$}& $C_4$ & $C_4$ & 1 & \multirow{4}{*}{1}\\
{} & $E_{8,4}$ & $E_{{4} \,x^iy^ix^jy^j,x^iy^ix^jy^j}$ &  3&\\
{} & $G_{10,6,2,2} $ & $G_{10,6,x^iy^i,x^iy^i}$ &  3&\\
{} & $I_{10,6,6,6}$ & $I_{10,6,6,6}$ & 1&\\
\hline
\rule[3 mm]{0 mm}{1 mm}\multirow{6}{*}{$({\bf 3}_{T_i} , {\bf 3}_{U_j}, {\bf 3}_{U_k}  )$}& $E_{8,4}$ & $E_{{4}\, x^iy^ix^jy^k,x^iy^ix^jy^k}$ &  2 &\multirow{6}{*}{3}\\
{} & {} & $E_{{4}\, x^iy^ix^jx^k,x^iy^ix^jx^k}$ & 1& \\
{} & {} & $E_{{4}\, x^iy^iy^jy^k,x^iy^iy^jy^k}$ & 1&\\
{} & $G_{10,6,2,2} $ & $G_{10,6,x^jy^k,x^jy^k}$&  2 &\\
{} & {} & $G_{10,6,x^jx^k,x^jx^k}$ & 1&\\
\rule[-2.5mm]{0mm}{2mm}
{} & {} & $G_{10,6,y^jy^k,y^jy^k}$ & 1& \\
\hline
\rule[3 mm]{0 mm}{1 mm}\multirow{8}{*}{$({\bf 3}_{T_i} , {\bf 3}_{U_i}, {\bf 3}_{U_j}  )$}& $E_{9,2,1}$ & $E_{{4}\, x^iy^ix^{j}y^{j}y^{k},x^iy^{k},x^i}$ & 1 & \multirow{8}{*}{6}\\
{} & {} & $E_{{4}\, x^iy^ix^{j}y^{j}y^{k},y^iy^{k},y^i}$ & 1 & \\
{} & {} & $E_{{4}\, x^iy^ix^{j}y^{j}x^{k},x^ix^{k},x^i}$ & 1 & \\
{} & {} & $E_{{4}\, x^iy^ix^{j}y^{j}x^{k},y^ix^{k},y^i}$ & 1 & \\
{} & $G_{10,5,4,1} $ & $G_{10,x^{j}y^{j}x^{k}y^{k}x^i,x^{j}y^{j}x^{k} x^{i},x^{k}}$ & 1 & \\
{} & {} & $G_{10,x^{j}y^{j}x^{k}y^{k}x^i,x^{j}y^{j}y^{k}x^{i},y^{k}}$ & 1 & \\
{} & {} & $G_{10,x^{j}y^{j}x^{k}y^{k}y^i,x^{j}y^{j}x^{k}y^{i},x^{k}}$ & 1 & \\
\rule[-2.5mm]{0mm}{2mm}
{} & {} & $G_{10,x^{j}y^{j}x^{k}y^{k}y^i,x^{j}y^{j}y^{k}y^{i},y^{k}}$ & 1 & \\
\hline
\rule[3 mm]{0 mm}{1 mm}\multirow{6}{*}{$({\bf 3}_{S} , {\bf 3}_{T_j}, {\bf 3}_{T_k}  )$} & $C_8$ & $C_{{4}\, x^jy^jx^ky^k}$& 1 & \multirow{6}{*}{3}\\
{} & $E_8 $ & $E_{{4}\, x^jy^jx^ky^k}$ & 1 & \\
{} & $E_{10,4,2} $ & $E_{10,x^iy^ix^jy^j,x^jy^j}$ & 2 &\\ 
{} & $G_{10,4,2} $ & $G_{10,x^iy^ix^jy^j,x^jy^j}$ & 2 &\\
{} & $G_{10,6,6,2} $ & $G_{10,6,6,x^iy^i}$ & 1&\\ 
\rule[-2.5mm]{0mm}{2mm}
{} & $I_{10,6,6,2} $ & $I_{10,6,6,x^iy^i}$ & 1&\\
\hline
\rule[3 mm]{0 mm}{1 mm}\multirow{6}{*}{$({\bf 3}_{S} , {\bf 3}_{U_j}, {\bf 3}_{U_k}  )$} & $E_{10,4,2}$ & $E_{10,x^jy^kx^iy^i,x^jy^k}$ &  2 & \multirow{6}{*}{3}\\
{} & {} & $E_{10,x^jx^kx^iy^i,x^jx^k}$ & 1&\\
{} & {} & $E_{10,y^jy^kx^iy^i,y^jy^k}$ & 1&\\
{} & $G_{10,4,2}$ & $G_{10,x^jy^kx^iy^i,x^jy^k}$ &  2&\\
{} & {} & $G_{10,x^jx^kx^iy^i,x^jx^k}$ & 1&\\
\rule[-2.5mm]{0mm}{2mm}
{} & {} & $G_{10,y^jy^kx^iy^i,y^jy^k}$ & 1&\\
\hline
\end{tabular}
}
\caption{\footnotesize{The $SL(2,\mathbb{R})^7$ representations of the IIB/O3 potentials which couple to the allowed space-filling branes, which have $\alpha$ odd. They are all the possible representations with four singlets and three triplets of $SL(2,\mathbb{R})|_{U}$. The branes are $2^{7}$ in number and organise in 16 different representations. \label{oddbranesB}}}
\end{center}
\end{table}
 
By looking at Tables \ref{CEbranestable} and \ref{Gbranestable}, one can notice that the potentials $E_{8,4}$ and $G_{10,6,2,2}$ couple also to additional branes that are not contained in the $({\bf 3}_{T_1} , {\bf 3}_{T_2}, {\bf 3}_{T_3}  )$ representation derived above. In particular, we are missing the components of these fields which transform under the action of the generators $\mathcal{U}_i$. Looking at $E_{8,4}$, one finds that for each $i$ there are four of those components, \textit{i.e.} $E_{{4}\, x^iy^ix^jy^k,x^iy^ix^jy^k}$ (two components for $i$ fixed), $E_{{4}\, x^iy^ix^jx^k,x^iy^ix^jx^k}$ and $E_{{4}\, x^iy^iy^jy^k,x^iy^iy^jy^k}$. One can also see that these components are singlets of $SL(2,\mathbb{R})|_{U_i}$  while transform with each other under the action of $\mathcal{U}_j$ and $\mathcal{U}_k$. Similarly, one can prove that they do not transform under the action of $\mathcal{T}_j$ and $\mathcal{T}_k$, whereas  under the action of $\mathcal{T}_i$  they generate the components $G_{10,6,x^jy^k,x^jy^k}$, $G_{10,6,x^jx^k,x^jx^k}$ and $G_{10,6,y^jy^k,y^jy^k}$ of  $G_{10,6,2,2}$. For fixed $i$, this gives a total of 8 branes, which collect in the representation that we denote $({\bf 3}_{T_i} , {\bf 3}_{U_j}, {\bf 3}_{U_k}  )$. The result is summarised in the second row of Table \ref{oddbranesB}.

The remaining $\alpha=-3$ and $\alpha=-5$ branes that are singlets under {S} are associated to the potentials $E_{9,2,1}$ and $G_{10,5,4,1}$. As Tables \ref{CEbranestable} and \ref{Gbranestable} show, this gives in total  48 branes, which 
are embedded in six different representations denoted as $({\bf 3}_{T_i} , {\bf 3}_{U_i}, {\bf 3}_{U_j}  )$. In particular, for fixed $i$ and $j$ 
the components $E_{{4}\, x^iy^ix^jy^jx^k,x^ix^k,x^i}$, $E_{{4}\, x^iy^ix^jy^jx^k,y^ix^k,y^i}$, $E_{{4}\, x^iy^ix^jy^jy^k,x^iy^k,x^i}$ and  $E_{{4}\, x^iy^ix^jy^jy^k,y^iy^k,y^i}$ of $E_{9,2,1}$, together with  the components $G_{10,x^jy^jx^ky^kx^i,x^jy^jx^ix^k,x^k}$, $G_{10,x^jy^jx^ky^kx^i,x^jy^jx^iy^k,y^k}$, $G_{10,x^jy^jx^ky^ky^i,x^jy^jy^iy^k,y^k}$ and $G_{10,x^jy^jx^ky^ky^i,x^jy^jy^ix^k,x^k}$ of $G_{10,5,4,1}$, belong to a single irreducible representation.

\begin{table}[t!]
\begin{center}
\begin{tabular}{|c|p{1.5 cm} p{4 cm}|c|c|}
\hline
\rule[3 mm]{0 mm}{2 mm}  IIA/O6 $SL(2,\mathbb{R})^7$ rep  & potential & component & \# & \# reps\\
\hline
\rule[3 mm]{0 mm}{2 mm}\multirow{4}{*}{$({\bf 3}_{U_1} , {\bf 3}_{U_2}, {\bf 3}_{U_3}  )$}& $C_7$ & $C_{{4}\, x^1x^2x^3}$ & 1& \multirow{4}{*}{1}\\
{} & $E_{9,3,1}$ & $E_{{4}\, x^iy^ix^jy^jx^k,y^iy^jx^k,x^k}$ & 3&\\
{} & $G_{10,5,3,1} $ & $G_{10,x^iy^ix^jy^jy^k,y^kx^ix^j,y^k}$ & 3&\\
\rule[-2.5mm]{0mm}{2mm}
{} & $I_{10,6,6,3}$ & $I_{10,6,6,y^1y^2y^3}$ & 1&\\
\hline
\rule[3 mm]{0 mm}{1 mm}\multirow{6}{*}{$({\bf 3}_{T_j} , {\bf 3}_{T_k}, {\bf 3}_{U_i}  )$}& $E_{9,3,1}$ & $E_{{4}\, x^iy^ix^ky^kx^j,y^ix^ky^k,x^k}$ &  2 & \multirow{6}{*}{3}\\
{} & $E_{8,1}$ & $E_{{4}\, x^iy^ix^jx^k,y^i}$ & 1&\\
{} & $E_{10,5,2}$ & $E_{10,y^ix^jy^jx^ky^k,x^jx^k}$ & 1&\\
{} & $G_{10,5,3,1}$ & $G_{10,x^iy^ix^ky^ky^j,x^ky^kx^i,y^k}$ & 1&\\
{} & $G_{10,6,5,2}$  & $G_{10,6,x^jx^ix^ky^jy^k,y^jy^k}$ &  2&\\
\rule[-2.5mm]{0mm}{2mm}
{} & $G_{10,4,1}$ & $G_{10,y^jy^kx^iy^i,x^i}$ & 1 &\\
\hline
\rule[3 mm]{0 mm}{1 mm}\multirow{8}{*}{$({\bf 3}_{T_i} , {\bf 3}_{T_j}, {\bf 3}_{U_i}  )$}& $E_{9,3,1}$ & $E_{{4}\,y^ix^{k}y^{k}x^{j}y^{j},x^{k}x^{j}y^{j},x^{j}}$ & 1& \multirow{8}{*}{6}\\
{} & {} & $E_{{4}\, x^iy^ix^{k}y^{k}x^{j},x^iy^ix^{k},y^i}$ & 1&\\
{} & $ E_{8,1}$ & $E_{{4}\, y^ix^{k}y^{k}x^{j},x^{k}}$ & 1&\\
{} & $ E_{10,5,2}$ & $E_{10,x^iy^ix^{k}x^{j}y^{j},y^ix^{j}}$ & 1&\\
{} & $G_{10,4,1} $ & $G_{10,x^{k}y^{k}y^{j}x^i,y^{k}}$ & 1&\\
{} & $G_{10,5,3,1} $ & $G_{10,x^{j}y^{j}x^{k}y^{k}x^i,x^{j}y^{j}y^{k},y^{j}}$ & 1&\\
{} & {} & $G_{10,x^iy^ix^{k}y^{k}y^{j},x^iy^iy^{k},x^i}$ & 1&\\
\rule[-2.5mm]{0mm}{2mm}
{} & $G_{10,6,5,2} $ & $G_{10,6,x^iy^ix^{j}y^{j}y^{k},x^iy^{j}}$ & 1&\\
\hline
\rule[3 mm]{0 mm}{1 mm}\multirow{6}{*}{$({\bf 3}_{S} , {\bf 3}_{U_j}, {\bf 3}_{U_k}  )$} & $C_7$ & $C_{{4}\, x^iy^jy^k}$  & 1 &\multirow{6}{*}{3}\\
{} & $E_{9,3,1} $ &  $E_{{4}\, x^jy^jx^{k}y^kx^i,x^jx^kx^i,x^i}$  & 1&\\
{} & {} & $E_{{4}\, y^jx^ky^kx^iy^i,x^ky^iy^j,y^j}$  & 2 &\\ 
{} & $G_{10,5,3,1} $ & $G_{10,x^iy^ix^jy^jx^k, y^jx^ix^k,x^k}$ &  2&\\
{} & {} & $G_{10,x^jy^jx^ky^ky^i,y^iy^jy^k,y^i}$ & 1&\\ 
\rule[-2.5mm]{0mm}{2mm}
{} & $I_{10,6,6,3} $ & $I_{10,6,6,x^jx^ky^i}$ & 1&\\
\hline
\rule[3 mm]{0 mm}{1 mm}\multirow{6}{*}{$({\bf 3}_{S} , {\bf 3}_{T_j}, {\bf 3}_{T_k}  )$} & $E_{9,3,1}$ & $E_{{4}\, x^iy^ix^ky^ky^j,x^ky^ky^i,y^k}$ & 2 &\multirow{6}{*}{3}\\
{} & $E_{10,5,2}$ & $E_{10,x^jy^jx^ky^ky^i,y^jy^k}$ & 1&\\
{} & $E_{8,1}$ & $E_{{4}\, x^iy^iy^jy^k,y^i}$ & 1&\\
{} & $G_{10,5,3,1}$ & $G_{10,x^jx^ky^kx^iy^i,x^ky^kx^i,x^k}$ &  2&\\
{} & $G_{10,4,1}$ & $G_{10,x^jx^kx^iy^i,x^i}$ & 1&\\
\rule[-2.5mm]{0mm}{2mm}
{} & $G_{10,6,5,2}$ & $G_{10,6,x^jy^jx^ky^kx^i,x^jx^k}$  & 1&\\
\hline
\end{tabular}
\caption{\footnotesize{The $SL(2,\mathbb{R})^7$ representations of the IIA/O6 potentials which couple to the allowed space-filling branes, which have $\alpha$ odd. They are all the possible representations with four singlets and three triplets of $SL(2,\mathbb{R})|_{T}$. The branes are $2^{7}$ in number and organise in 16 different representations.  \label{oddbranesA}}}
\end{center}
\end{table}

The D7-branes and their S-dual belong to triplets of $SL(2,\mathbb{R})|_{S}$. The allowed components of the electric potential of the D7-branes, $C_8$, are $C_{{4}\, x^iy^ix^jy^j}$, and one can see that they are invariant under the action of the three generators $\mathcal{U}_i$. Moreover, we find that these components do not transform under $\mathcal{T}_k$ while they generate respectively the components $E_{10,x^jy^jx^ky^k,x^jy^j}$ and $E_{10,x^iy^ix^ky^k,x^iy^i}$ of the potential $E_{10,4,2}$ under a single action of $\mathcal{T}_i$ and $\mathcal{T}_j$. By S-duality, we see that we have to include also the components $G_{10,x^jy^jx^ky^k,x^jy^j}$ and $G_{10,x^iy^ix^ky^k,x^iy^i}$ of $G_{10,4,2}$. From these components, under the composed action of the generators $\mathcal{T}_j$ and $\mathcal{T}_i$, we can generate the field $G_{10,6,6,x^ky^k}$ which forms a triplet of $SL(2,\mathbb{R})|_{S}$ together with its S-dual $I_{10,6,6,x^ky^k}$. The resulting representation is $({\bf 3}_{S} , {\bf 3}_{T_j}, {\bf 3}_{T_k}  )$.

\begin{table}[t!]
\begin{center}
\scalebox{1}{
\begin{tabular}{|c|p{1.5 cm} p{4.5 cm}|c|}
\hline
\rule[3 mm]{0 mm}{2 mm} IIB/O3 $SL(2,\mathbb{R})^7$ rep & potential & component & \# reps\\
\hline
\hline
\rule[3 mm]{0 mm}{2 mm}\multirow{8}{*}{$({\bf 3}_{T_j} , {\bf 3}_{T_k}, {\bf 3}_{U_k}  )$}  & $D_{7,1}$ & $D_{{4} \, x^{j}y^{j}x^{k},x^{k}}$& \multirow{8}{*}{6}\\
{} & {} & $D_{{4}\, x^{j}y^{j}y^{k},y^{k}}$ &\\
{} & $F_{10,4,1,1} $ & $F_{10,x^{k}y^{k}x^{i}y^{i},x^{k},x^{k}}$&\\
{} & {} & $F_{10,x^{k}y^{k}x^{i}y^{i},y^{k},y^{k}}$&\\
{} & $F_{9,5,2} $ & $F_{{4}\, x^i y^i x^{j} y^{j} x^{k},x^i y^i x^{j} y^{j}x^{k},x^{j}y^{j}}$& \\
{} & {} & $F_{{4}\, x^iy^ix^{j}y^{j}y^{k},x^iy^ix^{j}y^{j}y^{k},x^{j}y^{j}}$&\\
{} & $H_{10,6,5,3} $ & $H_{10,6,x^{j}y^{j}x^{i}y^{i}x^{k},x^{k}x^iy^i}$&\\
\rule[-2.5mm]{0mm}{2mm}
{} & {} & $H_{10,6,x^{j}y^{j}x^{i}y^{i}y^{k},y^{k}x^iy^i}$&\\
\hline
\rule[3 mm]{0 mm}{1 mm}\multirow{8}{*}{$({\bf 3}_{S} , {\bf 3}_{T_i}, {\bf 3}_{U_j}  )$} & $D_{9,3}$ & $D_{{4}\, x^{k}y^{k}x^iy^ix^{j},x^iy^ix^{j}}$ & \multirow{8}{*}{6}\\
{} & {} & $D_{{4}\, x^{k}y^{k}x^iy^iy^{j},x^iy^iy^{j}}$&\\
{} & $F_{9,3}$ & $F_{{4}\, x^{k}y^{k}x^iy^ix^{j},x^iy^ix^{j}}$&\\
{} & {} & $F_{{4}\, x^{k}y^{k}x^iy^iy^{j},x^iy^iy^{j}}$&\\
{} & $H_{10,6,3,1}$ & $H_{10,6,x^{k}y^{k}x^{j},x^{j}}$&\\
{} & {} & $H_{10,6,x^{k}y^{k}y^{j},y^{j}}$&\\
{} & $F_{10,6,3,1}$ & $F_{10,6,x^{k}y^{k}x^{j},x^{j}}$&\\
\rule[-2.5mm]{0mm}{2mm}
{} & {} & $F_{10,6,x^{k}y^{k}y^{j},y^{j}}$&\\
\hline
\end{tabular}
}
\caption{\footnotesize{The $SL(2,\mathbb{R})^7$ representations of the forbidden IIB/O3 potentials which couple to space-filling branes with $\alpha$ even. The branes are 96 in number and are embedded in 12 different representations. \label{evenbranesB}}}
\end{center}
\end{table}

Just as in the case of the S-duality singlets $E_{8,4}$ and $G_{10,6,2,2}$, that contain components that are not in the representation of the D3-branes, in this case there are components of the potentials $E_{10,4,2}$ and $G_{10,4,2}$ which do not transform in the same representation of the D7-branes; they are the components with the two extra indices labelling directions along different tori of the orbifold, which are listed in the last row of Table \ref{oddbranesB} and form the representation $({\bf 3}_{S} , {\bf 3}_{U_j}, {\bf 3}_{U_k}  )$.

To summarise, all the $SL(2,\mathbb{R})^{7}$ representations of the $2^{7}$ IIB/O3 branes have been collected in Table \ref{oddbranesB}. As it is clear by looking at the table, we find that the branes form 16 irreducible representations, each made of three triplets and four singlets, and therefore each containing 8 branes. The 16 representations are all the possible representations made of three triplets and four singlets and such that there is an even number of triplets with respect to   $SL(2,\mathbb{R})|_{U_i}$. By looking at Tables \ref{CEbranestable}, \ref{Gbranestable} and \ref{Ibranestable}, one can easily derive the mirror result for the IIA/O6 orientifold. This is listed in Table \ref{oddbranesA}. Obviously, in the IIA picture the representations are all those with an even number of triplets with respect to  $SL(2,\mathbb{R})|_{T_i}$.

In the previous section we have also derived the branes with even $\alpha$ that are forbidden by supersymmetry and whose absence implies Bianchi identities for the fluxes, as shown schematically in eq. \eqref{Bianchibranesintro}.
 Such branes are listed in Table \ref{alphaevenbranestable}. We repeat for these branes the same analysis carried out so far in this section for the branes with odd $\alpha$. As Table \ref{alphaevenbranestable} shows, there is a total of 96 such branes, and we find that  analogously to the branes with odd $\alpha$, they collect in 12 irreducible representations made of three triplets and four singlets. We give the full result in Table \ref{evenbranesB} for the IIB case and in Table \ref{evenbranesA} for the IIA case.

\begin{table}[t!]
\begin{center}
\begin{tabular}{|c|p{1.5 cm} p{4cm}|c|}
\hline
\rule[3 mm]{0 mm}{2 mm}IIA/O6 $SL(2,\mathbb{R})^7$ rep& potential & component &\# reps\\
\hline
\hline
\rule[3 mm]{0 mm}{2 mm}\multirow{8}{*}{$({\bf 3}_{T_k} , {\bf 3}_{U_j}, {\bf 3}_{U_k}  )$}  & $D_{7,1}$ & $D_{{4}\, x^{j}y^{j}x^i,x^i}$& \multirow{8}{*}{6}\\
{} & $D_{9,3}$ & $D_{{4}\, x^{j}y^{j}x^{k}y^{k}x^i,x^{k}y^{k}x^i}$&\\
{} & $F_{9,4,1} $ & $F_{{4}\, x^{j}y^{j}x^iy^iy^{k},x^{j}y^{k}x^iy^i,x^{j}}$&\\
{} & {} & $F_{{4}\, x^iy^ix^{j}y^{j}x^{k},x^iy^iy^{j}x^{k},y^{j}}$ &\\
{} & $F_{10,5,2,1} $ & $F_{10,x^{k}y^{k}x^{i}y^{i}x^{j},x^{j}y^{k},y^{k}}$&\\
{} & {} & $F_{10,x^iy^ix^{k}y^{k}y^{j},x^{k}y^{j},x^{k}}$ &\\
{} & $H_{10,6,5,3} $ & $H_{10,6,x^{j}y^{j}y^{i}x^{k}y^{k},x^{k}y^{k}y^i}$&\\
\rule[-2.5mm]{0mm}{2mm}
{} & $H_{10,6,3,1} $ & $H_{10,6,x^{j}y^{j}y^{i},y^i}$&\\
\hline
\rule[3 mm]{0 mm}{1 mm}\multirow{8}{*}{$({\bf 3}_{S} , {\bf 3}_{T_j}, {\bf 3}_{U_i}  )$} & $ D_{7,1}$ & $D_{{4}\, x^{k}y^{k}y^i,y^i}$& \multirow{8}{*}{6}\\
{} & $ D_{9,3}$ & $D_{{4}\, x^{k}y^{k}y^ix^{j}y^{j},y^ix^{j}y^{j}}$&\\
{} & $F_{9,4,1}$ & $F_{{4}\, x^{k}y^{k}x^iy^ix^{j},x^iy^ix^{j}x^{k},x^{k}}$& \\
{} & {} & $F_{{4}\, x^iy^ix^{k}y^{k}y^{j},y^{j}x^iy^iy^{k},y^{k}}$& \\
{} & $F_{10,5,2,1}$ & $F_{10,x^{j}y^{j}x^iy^ix^{k},x^{k}x^{j},x^{j}}$& \\
{} & {} & $F_{10,y^{k}x^iy^ix^{j}y^{j},y^{j}y^{k},y^{j}}$&\\
{} & $H_{10,6,3,1}$ & $H_{10,6,x^{k}y^{k}x^i,x^i}$&\\
\rule[-2.5mm]{0mm}{2mm}
{} & $H_{10,6,5,3}$ & $H_{10,6,x^{k}y^{k}x^{j}y^{j}x^i,x^ix^{j}y^{j}}$&\\
 \hline
\end{tabular}
\caption{\footnotesize{The $SL(2,\mathbb{R})^7$ representations of the forbidden IIA/O6 potentials which couple to space-filling branes with $\alpha$ even. The branes are 96 in number and are embedded in 12 different representations.\label{evenbranesA}}}
\end{center}
\end{table} 
 
We conclude this section by pointing out that the generators $\mathcal{T}_i$ and $\mathcal{U}_i$ can be written as the products of specific T-duality transformations and {S}. In particular, in IIB we find that 
\begin{align}
\nonumber &\mathcal{T}_{i} = - {\rm T}_{x^jy^jx^ky^k}{\rm S} {\rm T}_{x^jy^jx^ky^k}\\
&\mathcal{U}_i= - ({\rm S} {\rm T}_{x^iy^i})^3  \quad ,\label{sl2generators}
\end{align}
where ${\rm T}_{x^jy^jx^ky^k} = {\rm T}_{y^{k}} \circ {\rm T}_{x^{k}} \circ {\rm T}_{y^{j}} \circ  {\rm T}_{x^{j}}$ is the combination of single T-dualities along the directions $x^{j}$, $y^{j}$, $x^{k}$ and $y^{k}$, and similarly ${\rm T}_{x^iy^i} = {\rm T}_{y^{i}} \circ {\rm T}_{x^{i}}$ is the combination of T-dualities along $x^i$ and $y^i$. The reader can verify that using eq. \eqref{sl2generators} and the T-duality transformations of the fluxes given in eqs. \eqref{TdualityruleRRfluxes}, \eqref{TdualityruleNSfluxes}, \eqref{TdualityrulePfluxes}, \eqref{TdualityruleH'fluxes} and \eqref{TdualityruleF'fluxes} one recovers all the weights of the fluxes, and similarly using the universal T-duality rule given in eq. \eqref{allbranesallalphasruleintro} one can determine all the representations given in Tables \ref{oddbranesB} and \ref{evenbranesB}. Using mirror symmetry, an analogous result applies to the IIA case.

\section{Discussion and Conclusions}

The original motivation of this paper was to extend the analysis of \cite{Lombardo:2016swq}, where  the inclusion of the $\mathcal{P}$ fluxes was discussed for both the  IIB/O3 and IIA/O6 $T^6/[\mathbb{Z}_2 \times \mathbb{Z}_2 ]$ orientifold models,  to all the remaining allowed non-geometric fluxes. We have denoted such fluxes as NS$'$ and RR$'$, and we have shown that they transform under T-duality according to the general rules given in eqs. \eqref{TdualityruleH'fluxes} and \eqref{TdualityruleF'fluxes}. These rules allow to write down a complete expression for the superpotential, which is at most linear in the modulus $S$ and at most cubic in the moduli $T_i$ and $U_i$. The IIB and IIA expressions for the superpotential are related by mirror symmetry. 

The superpotential of the IIA/O6 theory given in eq. \eqref{superpotIIAfull} is obtained by  requiring the matching with IIB using the mapping dictated by the T-duality transformation rules that we have found. We interpret the fact that the IIA superpotential has the form $ \int e^{J_{\rm c}} \wedge ({\rm flux} \cdot \Omega_{\rm c}^n )$ as a general manifestation of mirror symmetry. Indeed, for any  $\mathcal{N}=1$ compactification with $SU(3)$ structure, mirror symmetry has been conjectured to correspond to the exchange of $\Omega$ and $e^{J_{\rm c}}$ \cite{Fidanza:2003zi,Grana:2005jc,Grana:2005tf,Grimm:2005fa,Grana:2006hr}, and if one assumes that this result is  actually reliable in a generic non-geometric setup, one can expect that the form of the IIA superpotential that we find is valid for any O6 orientifold. In this regard, it would be worth investigating our IIA result more deeply in the context of generalised geometry. 
Moreover, focusing on the large class of IIA/O6 Calabi-Yau orientifolds, it would be promising to understand the relation between the IIA superpotential and the cohomological numbers which characterise the topology of the compactification.  As far as we know, this relation has been worked out only for IIB/O3 Calabi-Yau orientifolds so far, including all the allowed RR, NS and $\mathcal{P}_1^{2}$ fluxes \cite{Blumenhagen:2015kja}.\footnote{The generalization of the Bianchi identities including $\mathcal{P}_1^{2}$ fluxes to the IIB/O3 Calabi-Yau orientifolds has been discussed as well  \cite{Shukla:2016hyy, Shukla:2016xdy}.}

We have also determined all the exotic branes that can be sourced by all the non-geometric fluxes. The tadpole conditions that we find have the general form given in eq. \eqref{generalformoftadpolesintro}. Similarly, we have determined how the inclusion of all the fluxes modifies the Bianchi identities which correspond to non-trivial quadratic constraints for the fluxes, arising in the orientifold from the absence of specific exotic branes of the maximal theory, as eq. \eqref{Bianchibranesintro} summarises. 
Exotic branes are charged with respect to mixed-symmetry potentials, and in particular denoting with $A_{p,q,r,..}$ a ten-dimensional mixed-symmetry potential, this corresponds to a brane if some of the  indices $p$  are isometries and contain all the indices $q$, which themselves contain all the indices $r$ and so on \cite{Bergshoeff:2011zk,Bergshoeff:2011mh,Bergshoeff:2011ee}. The crucial ingredient in our analysis of tadpole conditions and Bianchi identities is the
universal  rule discovered in \cite{Lombardo:2016swq} and given in eq. \eqref{allbranesallalphasruleintro}, which relates under T-duality the  brane components of the various mixed-symmetry potentials.

The mixed-symmetry potentials that occur in this paper arise in the decomposition of the Kac-Moody algebra $E_{11}$ \cite{West:2001as} with respect to the ten-dimensional theory \cite{Riccioni:2007au}, and in particular the brane components are associated to the real roots of $E_{11}$ \cite{Kleinschmidt:2011vu}. From the point of view of the lower-dimensional theory, these components  are the long weights of the representation of the symmetry group to which the corresponding potential belongs \cite{Bergshoeff:2013sxa}. The Bianchi identities that we find in section 3  are restricted to these components, as they are associated to branes  with even $\alpha$ that are forbidden by supersymmetry
(see eq. \eqref{Bianchibranesintro}). A more detailed analysis should include the Bianchi identities for the remaining components. We leave this as an open project.

An analogous restriction has actually been made on the fluxes themselves throughout the paper.
Indeed, we have always assumed that the fluxes that are not related by S and T-dualities to the RR and $\mathcal{H}_3$ fluxes  are not present. This implies that  $f_{ab}^a$ and $\mathcal{Q}_a^{ab}$ (with $a$ not summed) vanish, as well as all the $\mathcal{P}_a^{a b_1 ...b_p}$ fluxes. Similarly, for the $\mathcal{P}$, NS$'$ and RR$'$ fluxes with all upstairs indices, we only consider the components such that the first set of indices is fully contained  in the second. From a group-theoretic perspective,  all these components correspond again to the long weights of the $SO(6,6)$ representations in eq. \eqref{SO66representationsfluxes}. 

In \cite{Shelton:2006fd, Ihl:2007ah} it was shown that the Bianchi identities for the NS fluxes can be obtained via demanding the nilpotency of the generalized twisted differential operator $\mathcal{D}= \mathcal{H} \wedge +f \cdot + \mathcal{Q} \cdot +\mathcal{R} \cdot $, and it would be interesting to generalise this construction with all the fluxes included. 
We observe that many of the Bianchi identities and tadpole conditions that we have derived can be obtained by requiring independently the nilpotency of the operators $\mathcal{D}_1= ({\mathcal{H}}'^{3,6} \wedge+\mathcal{P}^{1,4} \cdot+\mathcal{Q} \cdot+ {\mathcal{F}}_3 \cdot)$ and its S-dual
$\mathcal{D}_2= ({\mathcal{F}}'^{3,6} \wedge+\mathcal{Q}'^{1,4} \cdot+\mathcal{P}_1^2 \cdot+ {\mathcal{H}}_3 \cdot)$.
The constraints obtained from the nilpotency of these operators have exactly the same form of the NS Bianchi identities when $\mathcal{P}$-fluxes are turned off (see, for instance, eq. (4.1) of \cite{Lombardo:2016swq}). In particular, from $(\mathcal{D}_1)^2=0$, we recover the Bianchi identities related to the fields $D_{9,3}$ and $F_{10,6,3,1}$ (see Table \ref{alphaevenbranestable}), and the tadpole conditions cancelled by the branes which couple to the fields $C_8$, $E_{10,4,2}$ and $G_{10,6,6,2}$. Similarly, the constraints worked out from $(\mathcal{D}_2)^2=0$ correspond to the Bianchi identities related to $F_{9,3}$ and $H_{10,6,3,1}$ and to the tadpoles cancelled by the branes coupling to $E_8$, $G_{10,4,2}$ and $I_{10,6,6,2}$.
It is interesting that also the Bianchi identities related to $D_{9,3}$ are in these constraints, thus overlapping with the whole set of NS Bianchi identities. This seems to suggest that a redundancy actually occurs when the nilpotencies of these operators are taken simultaneously into account. What is the real meaning of this, and whether there could exist an analogous description for the rest of the constraints discussed in section 3, remains to be investigated.

A natural continuation of our analysis would be a detailed study of the solutions to all the Bianchi identities and tadpole conditions that we have determined. Although this is beyond the scope of this paper, we have performed a preliminary numerical analysis aimed at finding solutions that are non-geometric and that can be allowed only in the presence of exotic branes. 
In the isotropic case, where the three two-dimensional tori are identified, we are left with $40$ flux parameters satisfying 16 Bianchi identities
and 30 tadpole conditions. We have looked for solutions by scanning the zero minima of the function defined as
the quadratic sum of all the constraints, where the integer number of branes and fluxes are generally treated as parameters to be determined. This was performed using the \textit{Mathematica} algorithm \textit{NMinimize}. 
As a sensible prescription, we discard the configurations of fluxes in which no minima are found in zero after a number of $10^5$ iterations. That is also justified by the fact that for any configuration of fluxes no new solutions are actually found in the range between $10^4$ and $10^5$ iterations.  

We give here a few IIB examples of the solutions we have found. If  only the fluxes $\mathcal{F}_3$, $\mathcal{H}_3$, $\mathcal{P}^{1,4}$ and $\mathcal{H}'^{3,6}$ are turned on, no solutions could be found without including exotic branes. Besides the D3-branes, the fluxes induce tadpoles for the exotic branes $3^4_3$, $6^{1,1}_{3}$, $5^{2,2}_3$  and $3^{0,4,2}_{5}$, and we find that the minimum number of $\alpha=-3$ branes which need to be included is equal to 3. If instead 
only the fluxes $\mathcal{F}_3$, $\mathcal{H}_3$, $\mathcal{H}'^{3,6}$ and $\mathcal{F}'^{3,6}$ are turned on, one finds that together with the D3-branes one has  to include the exotic branes $5^{2,2}_3$, $5^{2,2}_5$  and $3^{0,0,6}_7$. Interestingly, there are solutions every time at least one of the 3 different kinds of exotic branes is included. Finally, we have considered the case in which the fluxes $\mathcal{F}_3$, $\mathcal{H}_3$, $\mathcal{Q}'^{1,4}$ ,$\mathcal{H}'^{3,6}$ and $\mathcal{F}'^{3,6}$ are turned on.
This is the most constrained case that we have analysed (the total number of non-vanishing fluxes is 22), and  
to find solutions, besides the D3-branes, we need to include also the exotic branes $3^4_3$, $6^{1,1}_{3}$,$5^{2,2}_3$,$5^{2,2}_5$ and $3^{0,0,6}_7$. We plan to perform a more systematic search of solutions in the near future.

In \cite{Aldazabal:2006up} it has been shown that all the fluxes of the orientifold model fill the representation $(\mathbf{2},\mathbf{2},\mathbf{2},\mathbf{2},\mathbf{2},\mathbf{2},\mathbf{2})$ of $SL(2,\mathbb{R})^7$. Therefore, the fluxes are symmetric under the exchange of any pair of $SL(2,\mathbb{R})$'s.
In this paper we have shown that in IIB the space-filling branes group in 16 irreducible representations 
of $SL(2,\mathbb{R})^{7}$ which are all the possible representations made of four singlets and three triplets with an even number of triplets of  $SL(2,\mathbb{R})|_{U_i}$. 
This means that the symmetry under the exchange of the different $SL(2,\mathbb{R})$'s is  actually broken by the presence of space-filling branes. This reveals the real effect of the orientifold projection on the symmetries, which is to make a distinction between $SL(2,\mathbb{R})|_{U_i}$ and $SL(2,\mathbb{R})|_{T_i}$, whose exchange corresponds in fact to switching the description to IIA/O6. 
We have also expressed the action of the $SL(2,\mathbb{R})|_{U_i}$ and $SL(2,\mathbb{R})|_{T_i}$ generators in terms of S- and T-duality as in eq. \eqref{sl2generators}, which  is a general formula that can be applied to any representation. 
This on one side proves the power of our T-duality rules and on the other side gives an elegant expression for the $SL(2,\mathbb{R})^7$ generators which is universally valid.

To conclude, the inclusion of exotic branes may be an important tool in string compactifications with non-geometric fluxes. Indeed, it would enlarge the possibilities to find more viable $\mathcal{N}=1$ four-dimensional vacua where all moduli can be conveniently stabilised, as many constraints on the fluxes could be relaxed. 
Moreover, from a wider perspective which goes beyond the closed string sector, it could be of extreme interest to further extend  this analysis studying the dynamics of the exotic branes.

\vskip 1cm

\section*{Acknowledgements}
We would like to thank G. Dibitetto and G. Pradisi for discussions, and in particular R. D'Onofrio for applying the \textit{Mathematica} algorithm \textit{NMinimize} to derive various solutions of the tadpole conditions and Bianchi identities.
FR and SR would like to thank the University of Uppsala for hospitality during the final completion of this work.

\vskip 1.5cm


\end{document}